\documentclass[%
 reprint,
superscriptaddress,
 amsmath,amssymb,
 aps,
onecolumn
]{revtex4-2}

\usepackage{graphicx}
\usepackage{dcolumn}
\usepackage{bm}
\usepackage{xcolor, soul}
\usepackage{wrapfig}
\usepackage[T1]{fontenc}

\begin{document}

\preprint{APS/123-QED}

\title{Ion Transport in an Electrochemical Cell: A Theoretical Framework to Couple Dynamics of Double Layers and Redox Reactions for Multicomponent Electrolyte Solutions}
\author{Nathan Jarvey}
\author{Filipe Henrique}
\author{Ankur Gupta}
\email{ankur.gupta@colorado.edu}
\affiliation{%
Department of Chemical and Biological Engineering, University of Colorado, Boulder, CO }%

\date{\today}

\begin{abstract}
Electrochemical devices often consist of multicomponent electrolyte solutions. Two processes influence the overall dynamics of these devices: the formation of electrical double layers and chemical conversion due to redox reactions. However, due to the presence of multiple length and time scales, it is challenging to simulate both processes directly from the Poisson-Nernst-Planck equations. Therefore, common modeling approaches ignore one of the processes, assume the two are independent, or extrapolate the results from reaction-free systems.  To overcome these limitations, we formulate and derive an asymptotic model by solving the Poisson-Nernst-Planck equations for an arbitrary number of ions in the thin-double-layer limit. Our analysis reveals that there are two distinct timescales in the system: double-layer charging and bulk diffusion. Our model displays excellent quantitative agreement with direct numerical simulations. Further, our approach is computationally efficient and numerically stable, even for large potentials. We investigate the dynamics of charging for a binary electrolyte and three-ion system, and find that redox reactions impact the double-layer charging process at short times whereas they modify the double-layer capacitance at long times. Overall, the proposed theoretical framework advances our ability to simulate electrochemical devices that contain multiple ions and widens opportunities for future research in the field.  

\end{abstract}

\maketitle


\section{Introduction}
\label{sec: introduction}

\par{} Electrical double layers (EDLs) and reduction-oxidation (redox) reactions are commonly observed in electrochemical devices \cite{simon2010materials,newman2012electrochemical}. The charging of EDLs occurs due to electrostatic interactions, which results in a capacitive behavior between the electrode surface and the bulk solution \cite{simon2010materials}. Examples of electrochemical devices that rely on EDLs include supercapacitors \cite{becker1957} and capacitive deionizers \cite{HE2018282}, both of which employ porous materials to maximize surface area. While these systems rely on EDLs, redox reactions are commonly observed and are known to impact system dynamics \cite{he2016ageing}. In contrast to EDLs, redox reactions use a chemical mechanism to enable larger specific energy at the cost of reduced specific power \cite{simon2010materials}. Devices that primarily employ redox reactions include pseudocapacitors \cite{boota2019mxene} and batteries \cite{NITTA2015252, boota2019mxene}. Though these devices rely on redox reactions, a capacitive element is generally required to explain the dynamics observed in experiments \cite{schoetz2022disentangling}. In some applications, EDLs and redox reactions are employed simultaneously, such as in hybrid supercapacitors \cite{simon2010materials, simon2020perspectives} and Faradaic capacitive deionizers \cite{ZHANG2018314}. Recently, Gogotsi and Penner \cite{Gogotsi2018} noted that determining the relative contributions of EDLs and redox reactions to the overall current is vital to understanding the dynamics of electrochemical devices. As is evident from the  aforementioned discussion, the simultaneous inclusion of EDLs and redox reactions is required to accurately predict the dynamics of electrochemical devices. Yet, most continuum approaches that investigate charging dynamics ignore one of the mechanisms, as we detail below. Therefore, in this article, we propose a theoretical framework that enables us to investigate the simultaneous impact of EDLs and redox reactions on overall cell dynamics. 
\par{} We first discuss studies that are predicated on EDL charging. The majority of literature on EDLs is based on Gouy-Chapman-Stern (GCS) theory \cite{gouy1910constitution, chapman1913li, stern1924theorie}, which assumes that the EDL consists of two regions: the Stern layer and the diffuse layer \cite{deen2012analysis, lyklema1995fundamentals}.  The development of models that describe EDL charging has been pioneered  for porous materials by de Levie \cite{de1963porous,de1964porous} and for flat plate geometries by Bazant et al. \cite{bazant2004diffuse}. Specifically, Bazant et al. \cite{bazant2004diffuse} proposed a novel approach to solve the Poisson-Nernst-Planck (PNP) equations across all potentials in the limit of thin electrical double layers. This approach was expanded upon by Kilic and Bazant \cite{kilic2007steric,kilic2007stericpartone} to include the effects of finite ion size, as well as by Biesheuvel and Bazant \cite{biesheuvel2010nonlinear} for desalination applications in porous media. Feicht et al. \cite{feicht2016discharging} further developed a model that showcases the asymmetries in charging and discharging observed experimentally. Several studies also investigate the optimization of nanometer \cite{kondrat2013charging} or sub-nanometer \cite{kondrat2014accelerating} pores for supercapacitor applications. Additional effects such as the dielectric constant \cite{sawada2016introduction, gupta2018electrical}, electroconvection \cite{kim2019characterization}, Stefan-Maxwell fluxes \cite{balu2018role} and ion-ion correlations \cite{fedorov2008ionic,goodwin2017mean,Gonzalex-Tovar2004, gupta2018electrical} have also been studied, and can modify the overall dynamics of EDL charging \cite{bard2000electrochemical,newman2012electrochemical,rubinstein2000electro}. Recently, we developed models for porous media in the limits of overlapping \cite{gupta2020charging} and arbitrary \cite{henrique2022charging} double layers that  generalize the solution of the PNP equations in the small potential limit. While the reports mentioned above  advance models of EDL charging, they do not consider the presence of chemical reactions, which we discuss next. 
\par{} Separate from EDL modeling, the genesis of the kinetic modeling of heterogeneous electrochemical reactions occurring at or near electrode surfaces is commonly based upon the work of Butler \cite{butler1924studies,butler1924studies3, butler1932mechanism} and Volmer \cite{ErdeyGrazVolmer1930} through the development of the Butler-Volmer kinetic equation \cite{DICKINSON2020114145}. Later, Marcus theory \cite{marcus1951kinetics} and Marcus-Hush-Chidsey kinetics \cite{Kurchin2020,ZENG201477}  expanded upon Butler-Volmer kinetics  \cite{bard2000electrochemical,newman2012electrochemical,vetter2013electrochemical}. Recently, Rom\'{a}n et al. \cite{Roman2017} argued that for most electrochemical systems, these redox reaction kinetics can be captured effectively using Butler-Volmer kinetics. To model reactions near electrochemically active surfaces, the outer sphere approximation is commonly invoked, which states that surface redox reactions occur at the interface between the Stern and diffuse layers \cite{lyklema1995fundamentals}. Building upon previous work, Smith and Bazant \cite{smith2017multiphase} developed a porous electrode framework capable of numerically handling multiple electrode phases. Furthermore, several articles investigate the presence of vacancies and intercalation reactions in lithium ion battery electrodes while using Butler-Volmer kinetics to model the reactions occurring at the surface of and inside the electrodes  \cite{Sripad2020,liu2020interface,LATZ2013358,Boyle2020,Wu_2014}. However, these studies generally assume electroneutrality and thus overlook the dynamics of EDL charging. 
\par{} While the majority of electrochemical studies focus on either EDL charging or redox reactions, a limited set of reports investigates them simultaneously. For instance, Bazant et. al. \cite{bazant2005current,chu2005electrochemical} performed a steady state analysis for a flat plate geometry by combining the Butler-Volmer kinetic model with the solution of the PNP equations. Similarly, Biesheuvel et. al. \cite{biesheuvel2011diffuse} later developed a time-dependent solution for porous media, but crucially assumed the impacts of redox reactions and EDLs to be independent. More recent studies use density functional theory or molecular dynamics to investigate the effects of solvation shells \cite{Nakamura2011}, redox active end groups on overall capacitance \cite{su2020microscopic}, organic electrolytes for potassium-ion hybrid supercapacitors \cite{luo2019nonaqueous},  anion carriers within the electrolyte of zinc-ion hybrid supercapacitors \cite{Huangdft2021}, and proton transfer reactions at the surface of electrode materials on the kinetics of redox reactions within the EDL \cite{Dopke2022}. Additional models employ the solution of the PNP equations for carbon dioxide reduction environments \cite{bohra2019modeling}, asymmetric lithium ion hybrid supercapacitors \cite{campillo2019general}, a finite element method of solution for both symmetric and asymmetric valences \cite{PAZGARCIA2014263}, and grand canonical density functional theory methods \cite{alsunni2021electrocatalytic, sundararaman2017grand}.  
\par{}  In particular, we focus on the continuum approaches to simultaneously study the effect of EDLs and redox reactions \cite{bazant2005current,chu2005electrochemical, marcicki2014lithium,lsv2017}. One of the key challenges in such approaches is that to evaluate the potential profile in the bulk region, the potential drop across the EDLs needs to be determined a priori. To circumvent this complication, studies often employ charge and voltage relationships derived for EDLs without a chemical reaction \cite{biesheuvel2011diffuse, lsv2017,marcicki2014lithium}, which may not be applicable to systems with reactions. To make progress in resolving these discrepancies, we take inspiration from the work of Bazant et al. \cite{bazant2004diffuse}, who performed a perturbation analysis- commonly known as the BTA analysis- for EDL charging in a flat plate geometry for a binary symmetric electrolyte without any chemical reactions. In this article, we generalize the BTA analysis for an arbitrary number of ions and surface redox reactions. Our analysis reveals two distinct timescales- one for double-layer formation, and another for bulk diffusion- which have different relationships between electromigrative and diffusive fluxes. We derive separate boundary conditions for each timescale and also provide self-consistent expressions to evaluate concentrations and potentials within EDLs. Finally,  we asymptotically match the solutions in both space and time together to form one composite solution. 

\par{} We validate our model through direct numerical simulations for a range of physical parameters. The proposed mathematical framework shows that the EDL charging process is coupled with the redox reaction process, and they should not be treated independently. As an additional advantage, it is straightforward to numerically solve the derived set of equations, and  the computational cost is lower by an order of magnitude when compared to direct numerical simulations. Furthermore, our approach is not susceptible to numerical instabilities, which are often reported in numerical solutions of PNP equations and thus need to handled carefully \cite{mirzadeh2014conservative,mirzadeh2014enhanced,amrei2018oscillating}. Our work provides a theoretical framework to couple EDL charging and redox reactions and simulate electrochemical cells with minimal computational cost, enabling future studies to investigate these effects simultaneously. In this contribution, we do not include the effects of transport inside the electrodes, and the reader is referred to the work of Smith and Bazant for considerations related to multiphase ionic transport \cite{smith2017multiphase}.

\par{} We first describe the setup of the problem (section \ref{sec: formulation}), followed by a dimensionless formulation (section \ref{sec: dim-formulation}). We perform a perturbation analysis at two distinct timescales (sections \ref{sec: short}, \ref{sec: long}) and present a composite solution  (section \ref{sec: stitch}). We validate our results and describe the impact of coupled EDLs and redox reactions for a symmetric binary electrolyte with a constant flux (section \ref{sec: 2ion}). To demonstrate the universality of our approach, we also investigate the a three ion scenario (section \ref{sec: 3ion}). Finally, we summarize our results and provide suggestions for future directions (section \ref{sec: conclusions}).


\section{Problem Formulation}
\label{sec: formulation}
We consider a standard one-dimensional electrochemical cell with two electrodes separated by a distance $2 L$. The center of the cell is denoted by $X=0$ such that the electrodes are situated at $X=\pm L$; see Fig. \ref{Fig: schematic}. The region between the two electrodes is filled with an arbitrary number of ions of concentration $C_i (X,\tau)$, where $i$ denotes the $i^{\textrm{th}}$ ion and $\tau$ is time. The electric potential is denoted as $\Phi(X,\tau)$. The transport of ions is governed by Nernst-Planck equations 
\begin{subequations}
\label{Eq: dim-eqn}
\begin{equation}
\frac{\partial C_i}{\partial \tau}  + \frac{\partial N_i}{\partial X} = 0,  
\label{Eq: nernst-planck}
\end{equation}

\begin{wrapfigure}[40]{R}{0.5\textwidth}
    \centering
    \includegraphics[width = 0.45 \textwidth]{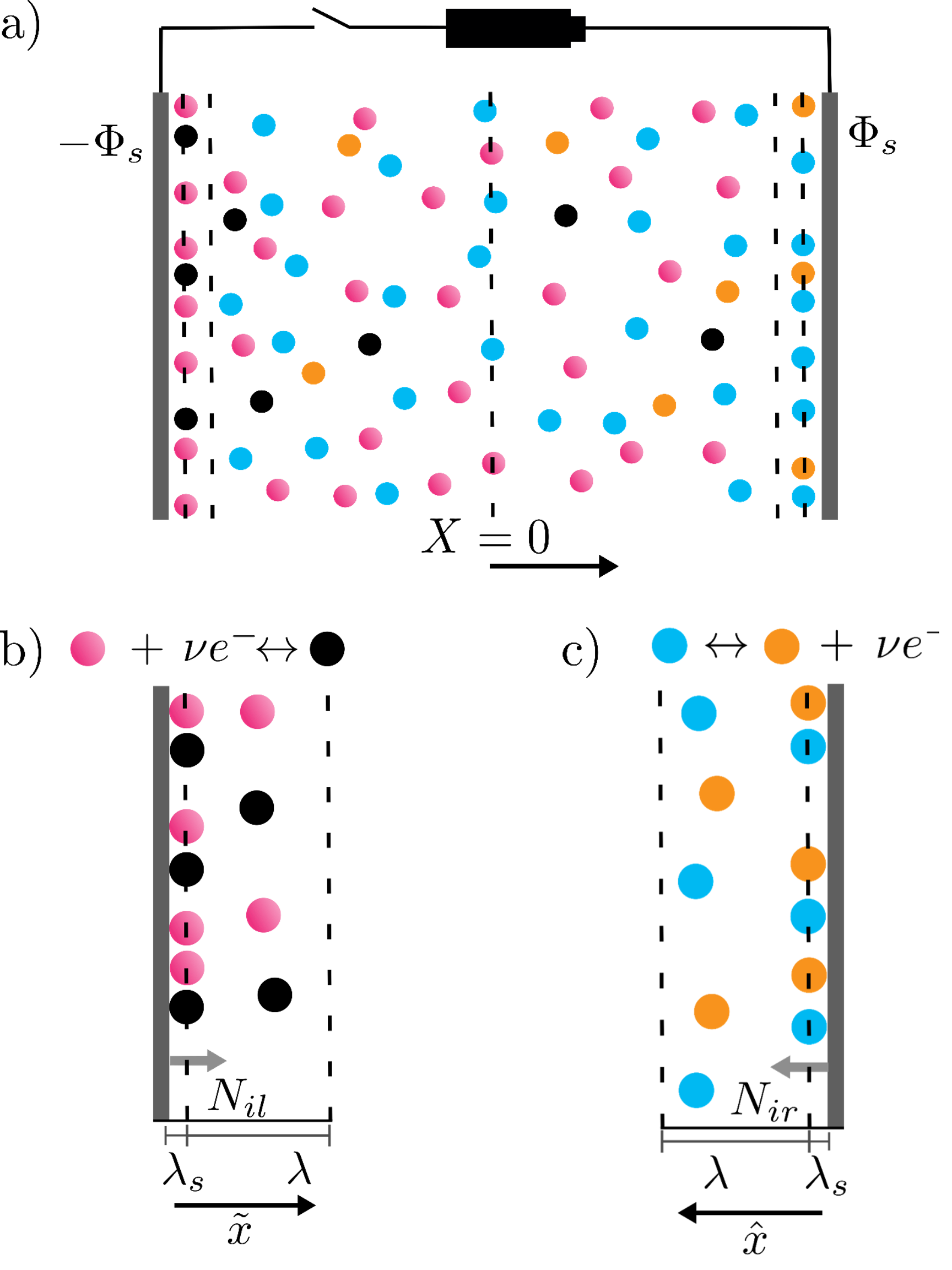}
    \caption{{\textbf{Schematic of the model problem.} A schematic of the model problem a) across the entire cell, b) in the left electrical double layer, c) in the right electrical double layer.  $\lambda_s$ is the Stern layer length, $\lambda$ is the diffuse layer length, and $2L$ is the total length of the system, with $X = 0$ at the centerline. The spatial coordinate $\tilde{x}$ is used within the left double layer, which has flux boundary condition $N_{il}$ for the $i^{\textrm{th}}$ ion, while coordinate $\hat{x}$ and boundary condition $N_{ir}$ are used within the right double layer. At $t = 0$, the switch is closed and the potentials at the left and right electrodes are $\mp\Phi_s$, respectively.}}
    \label{Fig: schematic}
\end{wrapfigure} 

where $N_i$ is the flux of the $i^{\textrm{th}}$ ion and is given by 
\begin{equation}
N_i = - \mathcal{D} \frac{\partial C_i}{ \partial X} - \frac{\mathcal{D} z_i e C_i}{k_B T} \frac {\partial \Phi}{\partial X}.     
\label{Eq: flux}
\end{equation}
Here, $z_i$ is the valence of the $i^{\textrm{th}}$ ion, $\mathcal{D}$ is the diffusivity of the ions, $k_B$ denotes the Boltzmann constant, $T$ is temperature, and $e$ is the charge of an electron. We assume the diffusivity of ions to be equal because varying ion diffusivities introduces new timescales in the system \cite{amrei2018oscillating, balaji2020diffusivities,richter2020acrepulsion,henrique2022asymmetry} and may lead to inconsistencies in our proposed perturbation expansion, as we detail later. Our analysis can be used as a good initial approximation for systems with asymmetric diffusivities that do not contain strong acids or bases, as many ion diffusivities are within an order of magnitude of one another (roughly $10^{-9}$ $\textrm{m}^2 \textrm{s}^{-1}$) \cite{velegol2016origins}. The first term in the right-hand side of Eq. (\ref{Eq: flux}) is the diffusive flux whereas the second term is the electromigrative flux. While Eq. (\ref{Eq: flux}) implicitly assumes that the ions are point charges, it is straightforward to extend our analysis to include the finite size of ions \cite{kilic2007steric, kilic2007stericpartone}. Further, we note that we have currently ignored volumetric reactions, which may be important for certain electrochemical systems.  We comment on the inclusion of volumetric reactions within our analysis in Section \ref{sec: conclusions}.
\par{} To couple species transport with electrical potential, we write down the Poisson equation, i.e., 
\begin{equation}
    \label{Eq: poisson}
    - \varepsilon \frac{\partial^2 \Phi}{\partial X^2} = Q_e,
\end{equation}
where $\varepsilon$ is the electrical permittivity of the solvent and $Q_e = \sum_i z_i e C_i $ is the volumetric charge density. We note that while the Eqs. (\ref{Eq: nernst-planck}) - (\ref{Eq: poisson}) are self-consistent, it is sometimes convenient to also invoke the charge conservation equation, i.e., 
\begin{equation}
    \label{Eq: charge-cons}
    \frac{\partial Q_e}{\partial \tau} + \frac{\partial J}{\partial X}=0,
\end{equation}
\end{subequations}

where $J=\sum_i z_i e N_i$ is the net charge flux (or current per unit area). We emphasize that Eq. (\ref{Eq: charge-cons}) is not an independent equation and can be derived by multiplying Eq. (\ref{Eq: nernst-planck}) with $e z_i$ and performing a summation over all ions. Next, we discuss the initial and boundary conditions.  

\par{} At $\tau=0$, the concentration of the $i^{th}$ ion is denoted by 
\begin{subequations}
\begin{equation}
\label{Eq: dim-ic-ci}
C_i(X,0)=C_{i0}
\end{equation} 
such that charge density $Q_e(X,0) = \sum z_i e C_{i0}=0$. At $\tau=0$, a potential of $\Phi(\pm L, 0)=\pm \Phi_s$ is applied at the two electrodes. Since $Q_e(X,0)=0$, Eq. (\ref{Eq: poisson}) implies that 
\begin{equation}
\label{Eq: dim-ic-phi}
\Phi(X,0^{+}) = \Phi_s \frac{X}{L}.   
\end{equation}
\label{Eq: dim-ic}
\end{subequations}

We also assume that a Stern layer of thickness $\lambda_s$ is present at $X=\pm L$. For simplicity, we assume that the Stern layer does not consist of any ions such that $Q_e=0$ inside the Stern layer \cite{lyklema1995fundamentals}. Therefore, the potential inside the Stern layer region is linear; see Eq. (\ref{Eq: poisson}). The region of solution will thus be $-L + \lambda_s \le X \le L- \lambda_s$. Assuming that $\varepsilon$ is constant throughout the electrolyte region, the electric fields can be equated at the Stern-diffuse layer interface, or 
\begin{subequations}
\label{Eq: dim-bc}
\begin{equation}
    \label{Eq: elec-match}
    \left. \frac{\partial \Phi}{\partial X} \right|_{X=\mp L \pm \lambda_s} = \frac{\Phi_s \pm  \left. \Phi \right|_{X=\mp L \pm \lambda_s}}{\lambda_s}.
\end{equation}
To model redox reactions at electrodes, we set the flux at the Stern-diffuse layer boundary, commonly known as the outer-sphere approximation \cite{newman2012electrochemical}, i.e.,
\begin{eqnarray}
\label{Eq: redox-match-1}
N_i(-L + \lambda_s,\tau)=N_{il},  \\
\label{Eq: redox-match-2} N_i(L - \lambda_s,\tau)=N_{ir},
\end{eqnarray}
\end{subequations}
\noindent where $l$ and $r$ in subscripts denote the left and right electrode, respectively. We note that $N_{il}$ and $N_{ir}$ can be functions of ion concentrations and potentials. Additionally, while the analysis presented here focuses on the outer-sphere approximation, it could be extended to other models as well. Eqs. (\ref{Eq: nernst-planck}) - (\ref{Eq: redox-match-2}) describe a well-posed set of non-linear differential equations that can be solved for given physical properties of ions $\mathcal{D}, z_i$, and $C_{i0}$, redox fluxes $N_{il}$ and $N_{ir}$, applied potential $\Phi_s$, Stern layer thickness $\lambda_s$, and process parameters $T, \varepsilon$, and $L$. For the convenience of the reader, a list of symbols is provided in Table \ref{tab:nomenclature}.   

\par{} The goal of this study is to calculate $C_i(X,\tau)$ and $\Phi(X,\tau)$ for the governing equations, initial conditions, and boundary conditions described above. To do so, we invoke the thin double-layer assumption, i.e., $\frac{\lambda}{L} \ll 1$. $\lambda$ is a measure of the diffuse layer length, and its mathematical description will be provided in the next section. The thin double-layer approximation is a common assumption in electrochemistry and is generally valid for planar geometries \cite{lyklema1995fundamentals,newman2012electrochemical}. Typically, $C_{i0} = O(1)$ M, which implies that $\lambda=O(1)$ nm. In contrast, $L \gtrsim O(10) \ \mu$m, making the assumption valid. We do note that this assumption may not hold for porous geometries \cite{gupta2020charging, henrique2022charging, biesheuvel2011diffuse}. 

\section{Dimensionless Analysis}
\label{sec: dim-formulation}
Given the large number of parameters in this system, we write dimensionless concentration $c_i=\frac{C_i}{C^{*}}$, potential $\phi = \frac{e \Phi}{k_B T}$, spatial coordinate $x = \frac{X}{L}$, time $t = \frac{\tau \mathcal{D}}{L^2}$, flux $n_i = \frac{N_i L}{\mathcal{D} C^{*}}$, current $j_i = \frac{J_i L}{e \mathcal{D} C^{*}}$, charge density $\rho_e = \frac{Q_e}{e C^*}$, $\delta = \frac{\lambda_s}{\lambda}$ and $\epsilon = \frac{\lambda}{L}$, where $\lambda=\sqrt{ \frac{\varepsilon k_B T}{e^2 C^*}}$ is a measure of double-layer thickness, and $C^*$ is reference concentration. The definition of $\lambda$ is based on $C^*$ to avoid dependence on $z_i$ and $C_{i0}$ while scaling the equations. Here, we propose that time $t$ consists of two timescales: the double-layer charging timescale $t_{\lambda} = \frac{t}{\epsilon} = \frac{\tau \mathcal{D}}{\lambda L}$ and the bulk diffusion timescale $t_{D} = t = \frac{\tau \mathcal{D}}{L^2}$. The solution at the $t_{\lambda}$ scale makes up the short timescale solution, whereas the $t_D$ scale sets the long timescale behavior. The complete solution is evaluated by stitching the two solutions. We emphasize that when $c_i(x,t=0)$ = $c_i(x,t_{\lambda}=0)$ and  $c_i(x,t \rightarrow \infty)=c_i(x,t_D \rightarrow \infty)$. Additionally, $c_i(x,t_{\lambda} \rightarrow  \infty)=c_i(x,t_D=0)$. Similar expressions are also valid for $\phi$. Therefore, it should be noted that even though the definition of $t_D$ is identical to $t$, $t=0$ represents the initial condition whereas $t_D=0$ represents the steady state solution at the short timescale. For the convenience of the reader, a list of dimensionless symbols is provided in Table \ref{tab:nomenclature}. 

\par{} Before proceeding with the solution at two different timescales, we also note that the analysis here will rely on singular perturbation in the parameter $\epsilon$. Therefore, for both the short and long timescales, the solution will be divided into three spatial parts. The solution in the bulk of the electrochemical cell is indicated as $\bar{x} = x$ such that $ -1 + \epsilon \delta \le \bar{x} \le 1 - \epsilon \delta $ (Fig. \ref{Fig: schematic}(a)). The left electrical double-layer region (Fig. \ref{Fig: schematic}(b)) is defined through $\tilde{x} = \frac{x+1 - \epsilon \delta }{\epsilon}$ such that $0 \le \tilde{x} \le \infty$. Similarly, the right electrical double-layer region (Fig. \ref{Fig: schematic}(c)) is defined as $\hat{x} = \frac{-x + 1 - \epsilon \delta}{\epsilon}$ such that $0 \le \hat{x} \le \infty$. For consistency, in the remainder of this work, variables with bar are reserved for the bulk region analysis. Similarly, the variables with tilde and hat are reserved for left and right double-layer regions, respectively.  

\par{} We now write down the perturbed variables in powers of $\epsilon$. First, we focus on the bulk region 
\begin{subequations}
\label{Eq: var-bar}
\begin{eqnarray}
\bar{c}_{i} = \bar{c}_{i}^0 + \epsilon \bar{c}_{i}^1 + O(\epsilon^2), \\
\bar{\rho}_e = \bar{\rho}_e^0 + \epsilon \bar{\rho}_e^1 + O(\epsilon^2), \\  
\bar{\phi} =  \bar{\phi}^0 + \epsilon \bar{\phi}^1 + O(\epsilon^2), \\
\bar{n}_i = \bar{n}_i^0 + \epsilon \bar{n}_i^1 + O(\epsilon^2) \label{Eq: var-bar-ni}, \\ 
\bar{j} = \bar{j}^0 + \epsilon \bar{j}^1 + O(\epsilon^2), 
\end{eqnarray}
\end{subequations}
\noindent where the superscript of 0 refers to the leading order solution and the superscript of 1 refers to the first-order correction. In this study, we will mostly focus on the leading order solution. However, as we show later, the first-order terms are sometimes required to evaluate the leading order solution. Proceeding in a similar manner for the left double-layer region, 
\begin{subequations}
\begin{eqnarray}
\tilde{c}_{i} = \tilde{c}_{i}^0 + \epsilon \tilde{c}_{i}^1 + O(\epsilon^2), \\
\tilde{\rho}_e = \tilde{\rho}_e^0 + \epsilon \tilde{\rho}_e^1 + O(\epsilon^2), \\
\tilde{\phi} =  \tilde{\phi}^0 + \epsilon \tilde{\phi}^1 + O(\epsilon^2), \\
\tilde{n}_i = \tilde{n}_i^0 + \epsilon \tilde{n}_i^1 + O(\epsilon^2), \label{Eq: var-tilde-ni}\\ 
\tilde{j} = \tilde{j}^0 + \epsilon \tilde{j}^1 + O(\epsilon^2), 
\end{eqnarray}
\label{Eq: var-tilde}
\end{subequations}
and the right double-layer region,
\begin{subequations}
\begin{eqnarray}
\hat{c}_{i} = \hat{c}_{i}^0 + \epsilon \hat{c}_{i}^1 + O(\epsilon^2), \\
\hat{\rho}_{e} = \hat{\rho}_{e}^0 + \epsilon \hat{\rho}_{e}^1 + O(\epsilon^2), \\
\hat{\phi} =  \hat{\phi}^0 + \epsilon \hat{\phi}^1 + O(\epsilon^2), \\
\hat{n}_i = \hat{n}_i^0 + \epsilon \hat{n}_i^1 + O(\epsilon^2), \\ 
\hat{j} = \hat{j}^0 + \epsilon \hat{j}^1 + O(\epsilon^2). 
\end{eqnarray}
\label{Eq: var-hat}
\end{subequations}
Therefore, the values of $c_i$ and $\phi$ will be evaluated by stitching six different solutions, i.e., three spatial solutions for each of the two timescales. Next, we describe the short and long timescale solutions.   
\section{Short Timescale Solution} 
\label{sec: short}
The short timescale formulation is based on using $t_{\lambda}$ as the dimensionless time scaling. Within the short timescale formulation, there are three spatial regions as described above. 
\subsection{Bulk region analysis}
We  first focus on the bulk region. The governing equations described in Eq. (\ref{Eq: dim-eqn}) take the form
\begin{subequations}
\begin{eqnarray}
\frac{\partial \bar{c}_i}{\partial t_{\lambda}} + \epsilon \frac{\partial \bar{n}_i}{\partial \bar{x}} = 0, \label{Eq: bar-ci} \\
\bar{n}_i = - \frac{\partial \bar{c}_i}{\partial \bar{x}} - z_i \bar{c}_i \frac{\partial \bar{\phi}}{\partial \bar{x}},  \label{Eq: bar-ni} \\ 
- \epsilon^2 \frac{\partial^2 \bar{\phi}}{\partial \bar{x}^2} = \bar{\rho}_e,  \label{Eq: bar-poisson} \\
\frac{\partial \bar{\rho}_e}{\partial t_{\lambda}} + \epsilon \frac{\partial \bar{j}}{\partial \bar{x}} = 0.  \label{Eq: bar-current} 
\end{eqnarray}
\label{Eq: bar-eqn}
\end{subequations}
After substituting the expressions described in Eq. (\ref{Eq: var-bar}), the leading order equation balance, i.e., $O(1)$,  of Eq. (\ref{Eq: bar-ci}) yields
\begin{equation}
\frac{\partial \bar{c}_{i}^0 }{\partial t_{\lambda}} = 0  \implies \bar{c}_i^0 = c_{i0}, 
\label{Eq: ci0}
\end{equation}
where we have utilized the initial condition in Eq. (\ref{Eq: dim-ic-ci}). Next, we substitute Eqs. (\ref{Eq: ci0}) and (\ref{Eq: var-bar}) in Eq. (\ref{Eq: bar-ni}), and after collecting the $O(1)$ terms, we get
\begin{subequations}
\begin{equation}
\bar{n}_{i}^0 = - z_i c_{i0} \frac{\partial \bar{\phi}^0}{\partial \bar{x}}, 
\label{Eq: ni0}
\end{equation}
which implies that 
\begin{equation}
\bar{j}^0 = \sum_i z_i \bar{n}_i^0 = - \sum_i z_i^2 c_{i0} \frac{\partial \bar{\phi}^0}{\partial \bar{x}}.  
\label{Eq: j0}
\end{equation}
\end{subequations}
\begin{subequations}
Proceeding in a similar fashion, substituting Eq. (\ref{Eq: var-bar}) in Eq. (\ref{Eq: bar-poisson}) yields the following for the $O(1)$ and $O(\epsilon)$ balances
\begin{eqnarray}
\bar{\rho}_e^0 = 0 \label{Eq: rho0}, \\ 
\bar{\rho}_e^1 = 0 \label{Eq: rho1}.
\end{eqnarray}
\end{subequations}                                                                                                   
\begin{subequations}
Next, we substitute Eq. (\ref{Eq: var-bar}) in Eq. (\ref{Eq: bar-current}). The $O(1)$ balance simply yields $\bar{\rho}_e^0=0$, which is identical to Eq. (\ref{Eq: rho0}). However, the $O(\epsilon)$ balance gives    
\begin{equation}
    \frac{\partial \bar{j}^0}{\partial \bar{x}} = 0  \implies      \frac{\partial^2 \bar{\phi}^0}{\partial \bar{x}^2} = 0, 
    \label{Eq: d2phidx2}
\end{equation} 
where we have utilized the expression in Eq. (\ref{Eq: j0}). Finally, Eq. (\ref{Eq: d2phidx2}) implies that 
\begin{equation}
    \bar{\phi}^0 = A(t_{\lambda}) \bar{x} + B(t_{\lambda}),
    \label{Eq: phi0}
\end{equation}
\end{subequations}
such that $A(t_{\lambda})$ and $B(t_{\lambda})$ are to be determined. Based on the initial conditions described in Eq. (\ref{Eq: dim-ic-phi}), $A(0)=\phi_s$ and $B(0)=0$. In summary, Eqs. (\ref{Eq: ci0}) and (\ref{Eq: phi0}) complete the description of $\bar{c}_i^0$ and $\bar{\phi}^0$, and we need to solve the electrical double-layer regions to evaluate $A(t_{\lambda})$ and $B(t_{\lambda})$. 
\subsection{Left electrical double layer analysis}
\par{} First, we focus on the left electrical double-layer region. By changing the variable from $\bar{x}$ to $\tilde{x}$, the governing equations become
\begin{subequations}
\label{Eq: tilde-eqn}
\begin{eqnarray}
\epsilon \frac{\partial \tilde{c}_i}{\partial t_{\lambda}} +  \frac{\partial \tilde{n}_i}{\partial \tilde{x}} = 0, \label{Eq: tilde-ci} \\
\tilde{n}_i = - \frac{\partial \tilde{c}_i}{\partial \tilde{x}} - z_i \tilde{c}_i \frac{\partial \tilde{\phi}}{\partial \tilde{x}},  \label{Eq: tilde-ni} \\ 
- \frac{\partial^2 \tilde{\phi}}{\partial \tilde{x}^2} = \tilde{\rho}_e,  \label{Eq: tilde-poisson} \\
\epsilon \frac{\partial \tilde{\rho}_e}{\partial t_{\lambda}} +  \frac{\partial \tilde{j}}{\partial \tilde{x}} = 0.  \label{Eq: tilde-current} 
\end{eqnarray}
\end{subequations}
Next, we focus on Eq. (\ref{Eq: tilde-ci}). After substituting the expressions from Eq. (\ref{Eq: var-tilde}), the $O(1)$ balance yields that $\tilde{n}_i^0$ is constant. To determine the value of the constant, we invoke the matching conditions between the left electrical double-layer region and the bulk region, i.e.,
\begin{subequations}
\begin{equation}
\left. \tilde{n}_i \right|_{\tilde{x} \rightarrow \infty} = \epsilon \left. \bar{n}_i \right|_{\bar{x}=-1 + \epsilon \delta},  
\label{Eq: tilde-ni-match}
\end{equation}
which physically means that the far away flux in the electrical double-layer region is the same as the boundary flux in the bulk region; see Fig. 1. We note that Eq. (\ref{Eq: tilde-ni-match}) has an additional $\epsilon$ factor in the right-hand side because the definition of $\tilde{n}_i$ is based on $\tilde{x}$, while the definition of $\bar{n}_i$ is based on $\bar{x}$. Next, we substitute the expansions for $\bar{n}_i$ and $\tilde{n}_i$ from Eqs. (\ref{Eq: var-bar-ni}) and (\ref{Eq: var-tilde-ni}), respectively, into Eq. (\ref{Eq: tilde-ni-match}) and collect $O(1)$ terms to get 
\begin{equation}
    \left. \tilde{n}_i^0 \right|_{\tilde{x} \rightarrow \infty} = 0 \implies \tilde{n}_i^0(\tilde{x}, t_{\lambda}) = 0.
    \label{Eq: tilde-ni-zero}
\end{equation}
\end{subequations}
Physically, Eq. (\ref{Eq: tilde-ni-zero}) implies that due to the thinness of the double layer, the diffusive and electromigrative fluxes have to balance each other at $O(1)$. Next, we place Eq. (\ref{Eq: tilde-ni-zero}) into Eq. (\ref{Eq: tilde-ni}) to write
\begin{subequations}
\begin{equation}
    -\frac{\partial \tilde{c}_i^0}{\partial \tilde{x}} - z_i \tilde{c}_i^0 \frac{\partial \tilde{\phi^0}}{\partial \tilde{x}} = 0 \implies \frac{\partial}{\partial \tilde{x}} \left( \tilde{c}_i^0 \exp(z_i \tilde{\phi^0})  \right) = 0. 
    \label{Eq: PBtilde-der}
\end{equation}
To make further progress, we write down the matching conditions for $\tilde{c}_i$ and $\tilde{\phi}$, i.e.,
\begin{eqnarray}
 \left. \tilde{c}_i \right|_{\tilde{x} \rightarrow \infty} =  \left. \bar{c}_i \right|_{\bar{x}=-1 + \epsilon \delta} \implies \left. \tilde{c}_i^0\right|_{\tilde{x} \rightarrow \infty} = c_{i0} \label{Eq: tilde-ci-match} \\
  \left. \tilde{\phi}_i \right|_{\tilde{x} \rightarrow \infty} =  \left. \bar{\phi}_i \right|_{\bar{x}=-1 + \epsilon \delta} \implies \left. \tilde{\phi}_i^0\right|_{\tilde{x} \rightarrow \infty} = -A + B.  \label{Eq: tilde-phi-match} 
\end{eqnarray}
Integrating Eq. (\ref{Eq: PBtilde-der}) and utilizing the values derived above, we arrive at the well-known Boltzmann distribution, i.e., 
\begin{equation}
    \tilde{c}_i^0 = c_{i0} \exp \left(-z_i \left(\tilde{\phi}^0 + A - B \right) \right).
    \label{Eq: tilde-PB}
\end{equation}
\end{subequations}
We emphasize that the derivation of Eq. (\ref{Eq: tilde-PB}) is valid regardless of the redox reactions at the Stern-diffuse layer interface. Therefore, the Boltzmann distribution holds at the double-layer charging timescale, which matches the results of Chamberlayne et. al. for weak electrolytes \cite{chamberlayne2020effects}. Next, we focus on Eq. (\ref{Eq: tilde-poisson}). We substitute Eqs. (\ref{Eq: var-tilde}) into Eq. (\ref{Eq: tilde-poisson}) and collect the $O(1)$ terms to write
\begin{subequations}
\begin{equation}
    -\frac{\partial^2 \tilde{\phi}^0}{\partial \tilde{x}^2} = \tilde{\rho}_e^0 = \sum_i z_i \tilde{c}_i^0 = \sum_i z_i c_{i0} \exp \left(-z_i \left(\tilde{\phi}^0 + A - B \right) \right),
    \label{Eq: tilde-Poisson-boltzmann}
\end{equation}
where we have also substituted the result from Eq. (\ref{Eq: tilde-PB}). To solve Eq. (\ref{Eq: tilde-Poisson-boltzmann}), we need two boundary conditions. As $\tilde{x} \rightarrow \infty$, the matching condition described in Eq. (\ref{Eq: tilde-phi-match}) serves as one of the boundary conditions. Next, we utilize the boundary condition in Eq. (\ref{Eq: elec-match}) to write 
\begin{equation}
    \left. \delta \frac{\partial \tilde{\phi}}{\partial \tilde{x}} \right|_{\tilde{x}=0} = \phi_s + \tilde{\phi},
    \label{eq: tilde-Stern-general}
\end{equation}
which at $O(1)$ becomes
\begin{equation}
    \delta \tilde{q}= \phi_s + \tilde{\phi}^0, 
    \label{Eq: tilde-stern-bc}
\end{equation}
where the definition of $\tilde{q}$ is provided later in Eq. (\ref{Eq: tilde-rhoe-dx}). To make analytical progress in solving Eq. (\ref{Eq: tilde-Poisson-boltzmann}), we multiply both sides by $2 \frac{\partial \tilde{\phi}^0}{\partial \tilde{x}}$ and integrate to write 
\begin{equation}
    \left(  \left. \frac{\partial \tilde{\phi}^0}{\partial \tilde{x}}^2 \right) \right|_{\tilde{x} \rightarrow \infty}^{\tilde{x}}  = 2 \sum_i \left. c_{i0} \exp \left(-z_i \left( \tilde{\phi}^0 + A - B \right) \right)  \right|_{-A + B}^{\tilde{\phi}^0},
\end{equation}
which after substituting $\left.\frac{\partial \tilde{\phi}^0}{\partial \tilde{x}} \right|_{\tilde{x} \rightarrow \infty}=0$ 
and $\tilde{\phi}^0 (\tilde{x} \rightarrow \infty, t_{\lambda}) = - A+B$ becomes
\begin{equation}
     \frac{\partial \tilde{\phi}^0}{\partial \tilde{x}} = \sqrt{2 \sum_i c_{i0} \left( \exp \left[ -z_i \left(\tilde{\phi}^0 + A - B \right) \right] - 1 \right) },
     \label{Eq: dphitildedx}
\end{equation}
\end{subequations}
where the positive sign in the square root is because of the physics of the problem, i.e., the potential gradient in the left double layer is expected to be positive with $\tilde{x}$. Eq. (\ref{Eq: dphitildedx}) can be numerically integrated with Eq. (\ref{Eq: tilde-stern-bc}) as the boundary condition. 
\par{} Next, we focus on Eq. (\ref{Eq: tilde-current}). After substituting the expansions from Eq. (\ref{Eq: var-tilde}), we find that the $O(1)$ balance can also be inferred from Eq. (\ref{Eq: tilde-ni-zero}), i.e., $\tilde{j}^0 (\tilde{x}, t_{\lambda}) = \sum_i z_i \tilde{n}_i^0 = 0$. Next, we focus on the $O(\epsilon)$ balance, which yields
\begin{subequations}
\begin{equation}
\frac{\partial \tilde{\rho}_e^0}{\partial t_{\lambda}} + \frac{\partial \tilde{j}^1}{\partial \tilde{x}} = 0.
\label{Eq: tilde-curr-eps}
\end{equation}
While the values of $\tilde{j}^1$ and $\tilde{\rho}_e^0$ are not known, one can integrate Eq. (\ref{Eq: tilde-curr-eps}) as follows:
\begin{equation}
    \frac{d}{d t_{\lambda}} \left( \int_0^{\infty} \tilde{\rho}_e^0  \ d\tilde{x} \right) + \left. \tilde{j}^1 \right|_{0}^{\infty} = 0.
    \label{Eq: tilde-curr-eps-integral}
\end{equation}
To make further progress, we now focus on the charge density integral. We integrate Eq. (\ref{Eq: tilde-Poisson-boltzmann}) with $\tilde{x}$ from 0 to $\infty$ to write
\begin{equation}
    \tilde{q} = \int_0^{\infty} \tilde{\rho}_e^0  \ d\tilde{x} = \left. \frac{\partial \tilde{\phi}^0}{\partial \tilde{x}} \right|_{\tilde{x}=0} = \sqrt{2 \sum_i c_{i0} \left( \exp \left[ -z_i \left(\tilde{\phi}^0 (0) + A - B \right) \right] - 1 \right) },
    \label{Eq: tilde-rhoe-dx}
\end{equation}
where we have also made use of Eq. (\ref{Eq: dphitildedx}). Physically, $\tilde{q}$ refers to the total charge stored in the left EDL. To evaluate $\tilde{j}^1$ at the boundaries, we require the relationship given by the matching condition, i.e., $\tilde{j}(\infty, t_{\lambda}) = \epsilon \bar{j}(-1 + \epsilon \delta, t_{\lambda})$, where the factor of $\epsilon$ is due to the difference in the definitions of $\tilde{x}$ and $\bar{x}$. Therefore, 
\begin{equation}
    \tilde{j}^1(\infty, t_{\lambda}) = \bar{j}^0(-1 + \epsilon \delta, t_{\lambda}) = \sum_i z_i \bar{n}_i^0 = - \sum_i z_i^2 c_{i0} A,   
    \label{Eq: tilde-j1-infty}
\end{equation} 
where we have used Eqs. (\ref{Eq: j0}) and (\ref{Eq: d2phidx2}). We note that $\tilde{j}^1(0, t_{\lambda})$ can be evaluated based on the redox fluxes. Since $\tilde{j}^0(\tilde{x}, t_{\lambda})=0$ from the $O(1)$ balance, the redox fluxes impact the $O(\epsilon)$ scale. Therefore,
\begin{equation}
    \tilde{j}^1(0, t_{\lambda}) = \sum_i z_i n_{il} = j_l, 
    \label{Eq: tilde-j1-zero}
\end{equation}
where $j_l$ is the redox current at the left electrode and can be dependent on concentration and potential. Finally, to combine all the results, we define
\begin{equation}
\tilde{\gamma}(t_{\lambda}) = \tilde{\phi}^0(0, t_{\lambda}) + A(t_{\lambda}) - B(t_{\lambda}).   
\label{Eq: tilde-gamma}
\end{equation}
\end{subequations}
We combine Eqs. (\ref{Eq: tilde-stern-bc}), (\ref{Eq: dphitildedx}), and (\ref{Eq: tilde-gamma}) to write
\begin{subequations}
\begin{equation}
   \delta \tilde{q} = \phi_s + \tilde{\gamma} - A + B. 
   \label{Eq: tilde-stern-bc-final}
\end{equation}
We substitute Eqs. (\ref{Eq: tilde-rhoe-dx}), (\ref{Eq: tilde-j1-infty}), (\ref{Eq: tilde-j1-zero}), and (\ref{Eq: tilde-gamma}) in Eq. (\ref{Eq: tilde-curr-eps-integral}) to get
\begin{equation}
      \frac{d \tilde{q}}{d t_\lambda} = \sum_i z_i^2 c_{i0} A + j_l. 
      \label{Eq: tilde-curr-eqn-final}
\end{equation}
\label{Eq: tilde-final-Eqs}
\end{subequations}
Eqs. (\ref{Eq: tilde-final-Eqs}) are coupled ODE and algebraic equations with three variables $\tilde{\gamma}$, $A$, and $B$, but with two equations. Next, we also detail the right electrical double-layer region analysis to complete the system of equations. 
\subsection{Right electrical double layer analysis}

By changing the spatial variable from $\bar{x}$ to $\hat{x}$, the governing equations become
\begin{subequations}
\label{Eq: hat-eqn}
\begin{eqnarray}
\epsilon \frac{\partial \hat{c}_i}{\partial t_{\lambda}} +  \frac{\partial \hat{n}_i}{\partial \hat{x}} = 0, \label{Eq: hat-ci} \\
\hat{n}_i = - \frac{\partial \hat{c}_i}{\partial \hat{x}} - z_i \hat{c}_i \frac{\partial \hat{\phi}}{\partial \hat{x}},  \label{Eq: hat-ni} \\ 
- \frac{\partial^2 \hat{\phi}}{\partial \hat{x}^2} = \hat{\rho}_e,  \label{Eq: hat-poisson} \\
\epsilon \frac{\partial \hat{\rho}_e}{\partial t_{\lambda}} +  \frac{\partial \hat{j}}{\partial \hat{x}} = 0.  \label{Eq: hat-current} 
\end{eqnarray}
\end{subequations}
\noindent Employing a process similar to that of the left electrical double-layer region and substituting the expressions from Eq. (\ref{Eq: var-hat}), the $O(1)$ balance for Eq. (\ref{Eq: hat-ci}) illustrates that $\hat{n}_i^0$ is constant. Next, we apply the matching condition between the right electrical double-layer region and the bulk region, i.e.,
\begin{subequations}
\begin{equation}
\left. \hat{n}_i \right|_{\hat{x} \rightarrow \infty} = - \epsilon \left. \bar{n}_i \right|_{\bar{x}=1-\epsilon \delta}.  
\label{Eq: hat-ni-match}
\end{equation}
By collecting $O(1)$ terms from Eq. (\ref{Eq: hat-ni-match}), we reach
\begin{equation}
    \left. \hat{n}_i^0 \right|_{\hat{x} \rightarrow \infty} = 0 \implies \hat{n}_i^0(\hat{x}) = 0.
    \label{Eq: hat-ni-zero}
\end{equation}
\end{subequations}
Similar to the left electrical double-layer region, the $O(1)$ balance for Eq. (\ref{Eq: hat-ni}) combined with Eq. (\ref{Eq: hat-ni-zero}) and the matching condition for $\hat{c}_i$ and $\hat{\phi}$ yields the Boltzmann distribution, i.e.,
\begin{equation}
    \hat{c}_i^0 = c_{i0} \exp \left(-z_i \left(\hat{\phi}^0  - A - B \right) \right).
    \label{Eq: hat-PB}
\end{equation}
We substitute Eq. (\ref{Eq: hat-PB}) in Eq. (\ref{Eq: hat-poisson}) and collect the $O(1)$ terms to write
\begin{subequations}
\begin{equation}
    -\frac{\partial^2 \hat{\phi}^0}{\partial \hat{x}^2} = \hat{\rho}_e^0 = \sum_i z_i \hat{c}_i^0 = \sum_i z_i c_{i0} \exp \left(-z_i \left(\hat{\phi}^0 - A - B \right) \right),
    \label{Eq: hat-Poisson-boltzmann}
\end{equation}
where from the matching condition $\hat{\phi} (\hat{x} \rightarrow \infty, t_{\lambda}) = A(t_{\lambda})+B (t_{\lambda})$. Application of the boundary condition in Eq. (\ref{Eq: elec-match}) and collecting $O(1)$ terms reveals 
\begin{equation}
    \delta \hat{q} = \hat{\phi}^0 - \phi_s, 
    \label{Eq: hat-stern-bc}
\end{equation}
where the definition of $\hat{q}$ is provided later in Eq. (\ref{Eq: hat-rhoe-dx}). After multiplying both sides of Eq. (\ref{Eq: hat-Poisson-boltzmann}) by $2 \frac{\partial \hat{\phi}^0}{\partial \hat{x}}$, integrating, and simplifying like the left electrical double layer, we get 
\begin{equation}
     \frac{\partial \hat{\phi}^0}{\partial \hat{x}} = - \sqrt{2 \sum_i c_{i0} \left( \exp \left[ -z_i \left(\hat{\phi}^0 - A - B \right) \right] - 1 \right) },
     \label{Eq: dphihatdx}
\end{equation}
\end{subequations}
where the negative sign comes as a result of the physics of the problem, i.e., the potential gradient in the right double layer is expected to be negative with $\hat{x}$ due to the direction of $\hat{x}$. 
\par{} For Eq. (\ref{Eq: hat-current}), an $O(1)$ balance yields $\hat{j}^0 = 0$. The $O(\epsilon)$ balance integrated from $0 \le \hat{x} \le \infty$ elucidates
\begin{subequations}

\begin{equation}
    \frac{d}{d t_{\lambda}} \left( \int_0^{\infty} \hat{\rho}_e^0  \ d\hat{x} \right) + \left. \hat{j}^1 \right|_{0}^{\infty} = 0.
    \label{Eq: hat-curr-eps-integral}
\end{equation}
The charge density integral evaluated by using Eq. (\ref{Eq: hat-Poisson-boltzmann}) becomes \begin{equation}
   \hat{q} = \int_0^{\infty} \hat{\rho}_e^0  \ d\hat{x} = \left. \frac{\partial \hat{\phi}^0}{\partial \hat{x}} \right|_{\hat{x}=0} = - \sqrt{2 \sum_i c_{i0} \left( \exp \left[ -z_i \left(\hat{\phi}^0 (0) - A - B \right) \right] - 1 \right) }.
    \label{Eq: hat-rhoe-dx}
\end{equation}
Physically, $\hat{q}$ refers to the total charge stored in the right EDL. To evaluate $\hat{j}^1$ at the boundaries, we note that from the matching condition, $\hat{j}(\infty, t_{\lambda}) = - \epsilon \bar{j}(1 - \epsilon \delta, t_{\lambda})$. Therefore, 
\begin{equation}
    \hat{j}^1(\infty, t_{\lambda}) = - \bar{j}^0(1 - \epsilon \delta, t_{\lambda}) = - \sum_i z_i \bar{n}_i^0 =  \sum_i z_i^2 c_{i0} A,   
    \label{Eq: hat-j1-infty}
\end{equation} 
where we have used Eqs. (\ref{Eq: j0}) and (\ref{Eq: d2phidx2}). Similarly,
\begin{equation}
    - \hat{j}^1(0, t_{\lambda}) =  \sum_i z_i n_{ir} =  j_r, 
    \label{Eq: hat-j1-zero}
\end{equation}
where $j_r$ is the redox current at the right electrode. We define
\begin{equation}
\hat{\gamma}(t_{\lambda}) = \hat{\phi}^0(0,t_{\lambda}) - A(t_{\lambda}) - B(t_{\lambda}).   
\label{Eq: hat-gamma}
\end{equation}
\end{subequations}
We combine Eqs. (\ref{Eq: hat-stern-bc}), (\ref{Eq: dphihatdx}), and (\ref{Eq: hat-gamma}) to get
\begin{subequations}
\begin{equation}
   -\delta \hat{q} = \phi_s - \hat{\gamma} - A - B,
   \label{Eq: hat-stern-bc-final}
\end{equation}
and we substitute Eqs. (\ref{Eq: hat-rhoe-dx}), (\ref{Eq: hat-j1-infty}), (\ref{Eq: hat-j1-zero}), and (\ref{Eq: hat-gamma}) in Eq. (\ref{Eq: hat-curr-eps-integral}) to get
\begin{equation}
      \frac{d \hat{q}}{d t_\lambda}= - \sum_i z_i^2 c_{i0} A - j_r. 
      \label{Eq: hat-curr-eqn-final}
\end{equation}
\label{Eq: hat-final-Eqs}
\end{subequations}
\subsection{Combining the three regions}
Eqs. (\ref{Eq: tilde-rhoe-dx}), (\ref{Eq: tilde-final-Eqs}), (\ref{Eq: hat-rhoe-dx}), and (\ref{Eq: hat-final-Eqs}) complete the system of equations required for the short timescale. These equations can be solved simultaneously for $\tilde{\gamma}$, $\hat{\gamma}$, $\tilde{q}$, $\hat{q}$, $A$, and $B$ with initial conditions $\tilde{\gamma}(0)=0$, $\hat{\gamma}(0)=0$, $\tilde{q}(0)=0$, $\hat{q}(0)=0$, $A(0)=\phi_s$, and $B(0)=0$. Next, we can evaluate $\tilde{\phi}^0 (\tilde{x}, t_{\lambda})$ and $\hat{\phi}^0 (\hat{x}, t_{\lambda})$ by integrating Eqs. (\ref{Eq: dphitildedx}) and (\ref{Eq: dphihatdx}) for each time step.
Finally, the stitched solutions read
\begin{subequations}
\begin{equation}
\begin{aligned}
c_i^{\lambda} (x, t_{\lambda}) = c_{i0} &+ c_{i0}  \left( \exp \left[- z_i \left\{  \tilde{\phi}^0 \left( \frac{x + 1 - \epsilon \delta }{\epsilon}, t_{\lambda}  \right) + A(t_{\lambda}) - B(t_{\lambda}) \right \} \right] -1 \right)  \\ 
     &+ c_{i0}  \left( \exp \left[- z_i \left\{  \hat{\phi}^0 \left( \frac{x - 1 + \epsilon \delta}{\epsilon}, t_{\lambda} \right) - A(t_{\lambda}) - B(t_{\lambda}) \right \} \right] -1 \right), 
\end{aligned}  
\end{equation}
\begin{equation}
    \phi^{\lambda}(x,t_{\lambda}) = A(t_{\lambda}) x - B(t_{\lambda}) + \tilde{\phi}^0  \left( \frac{x + 1 - \epsilon \delta }{\epsilon}, t_{\lambda}  \right) + \hat{\phi}^0 \left( \frac{x - 1 + \epsilon \delta}{\epsilon}, t_{\lambda} \right). 
\end{equation}
\label{Eq: lamda-full-sol}
\end{subequations}

\section{Long Timescale Solution}
\label{sec: long}
We now derive the results for the three regions at the long timescale, i.e., $t_D$. As with the short timescale, we initially describe the bulk region, followed by the left and right electrical double-layer regions. 
\subsection{Bulk region analysis}
The governing equations with the $t_D$ scale and $\bar{x}$ take the form
\begin{subequations}
\begin{eqnarray}
\frac{\partial \bar{c}_i}{\partial t_D} +  \frac{\partial \bar{n}_i}{\partial \bar{x}} = 0, \label{Eq: bar-ci-tD}\\
\bar{n}_i = - \frac{\partial \bar{c}_i}{\partial \bar{x}} - z_i \bar{c}_i \frac{\partial \bar{\phi}}{\partial \bar{x}} \label{Eq: bar-ni-tD} \\
- \epsilon^2 \frac{\partial^2 \bar{\phi}}{\partial \bar{x}^2} = \bar{\rho}_e,  \label{Eq: bar-poisson-tD} \\
\frac{\partial \bar{\rho}_e}{\partial t_D} +  \frac{\partial \bar{j}}{\partial \bar{x}} = 0 \label{Eq: bar-current-tD}. 
\end{eqnarray}
\end{subequations}

Eqs. (\ref{Eq: bar-ci-tD}) and (\ref{Eq: bar-ni-tD}) at $O(1)$ simply yield
\begin{equation}
    \frac{\partial \bar{c}_i^0}{\partial t_D} +  \frac{\partial \bar{n}_i^0}{\partial \bar{x}} = 0, 
 \label{Eq: bar-ci0-tD} 
\end{equation}
and
\begin{equation}
    \bar{n}_i^0 = -\frac{\partial \bar{c}_i^0}{\partial \bar{x}} - z_i  \bar{c}_i^0 \frac{\partial \bar{\phi}^0}{\partial \bar{x}}
    \label{Eq: bar-ni0-tD}
\end{equation}
Eq. (\ref{Eq: bar-poisson-tD}) at $O(1)$ and $O(\epsilon)$ yields $\bar{\rho}_e^0=0$ and $\bar{\rho}_e^1=0$. Using this in Eq. (\ref{Eq: bar-current-tD}) at $O(1)$ reveals
\begin{equation}
    \frac{\partial \bar{j}^0}{\partial \bar{x}} =0 \implies \bar{j}^0=c_1 \implies -\sum_i z_{i}^2 c_{i0} \frac{\partial \bar{\phi}^0}{\partial \bar{x}} = c_1,
    \label{Eq: bar-j0-tD}
\end{equation}
where $c_1$ is to be determined. Eqs. (\ref{Eq: bar-ci0-tD}), (\ref{Eq: bar-ni0-tD}) and (\ref{Eq: bar-j0-tD}) yield a consistent set of equations to solve for $\bar{c}_i^0 (\bar{x},t_D)$ and $\bar{\phi}^0(\bar{x},t_D)$. The initial condition in $t_D$ can be obtained via the steady state solution at the $t_{\lambda}$ scale, i.e., $\bar{c}_i^0 (\bar{x},0)=c_{i0}$. Because there is no time derivative associated with $\bar{\phi}^0$ in Eq. (\ref{Eq: bar-j0-tD}), no initial condition is required for $\bar{\phi}^0(\bar{x}, t_D)$. For a self-consistent formulation, we do require the initial $\bar{\phi}^0(\bar{x}, t_D=0)$ to agree with the steady state condition at the $t_\lambda$ scale, or $\bar{\phi}^0(\bar{x}, t_\lambda \rightarrow \infty)$. This requirement holds only when ion diffusivities are equal, which is why the analysis presented in this paper is restricted to the case of equal diffusivities. Next, to obtain the boundary conditions, we focus on the electrical double-layer regions.
\subsection{Left electrical double-layer region}
For the left electrical double-layer region, the governing equations at the $t_D$ timescale read
\begin{subequations}
\begin{eqnarray}
\epsilon^2 \frac{\partial \tilde{c}_i}{\partial t_D} +  \frac{\partial \tilde{n}_i}{\partial \tilde{x}} = 0, \label{Eq: tilde-ci-tD}\\
\tilde{n}_i = - \frac{\partial \tilde{c}_i}{\partial \tilde{x}} - z_i  \tilde{c}_i \frac{\partial \tilde{\phi}}{\partial \tilde{x}} \label{Eq: tilde-ni-tD} \\
- \frac{\partial^2 \tilde{\phi}}{\partial \tilde{x}^2} = \tilde{\rho}_e,  \label{Eq: tilde-poisson-tD} \\
\epsilon^2 \frac{\partial \tilde{\rho}_e}{\partial t_D} +  \frac{\partial \tilde{j}}{\partial \tilde{x}} = 0 \label{Eq: tilde-current-tD}. 
\end{eqnarray}
\end{subequations}
Proceeding similar to the $t_{\lambda}$ scale, $O(1)$ balance of Eq. (\ref{Eq: tilde-ci-tD}) illustrates that $\tilde{n}_i^0$ and $\tilde{n}_i^1$ are constant in $\tilde{x}$. Using these along with the matching condition in Eq. (\ref{Eq: tilde-ni-match}) and the redox flux equation, $O(1)$ and $O(\epsilon)$ balances show that
\begin{subequations}
\begin{eqnarray}
 \tilde{n}_i^0 = 0 \label{Eq: tilde-ni0-tD}, \\
 \tilde{n}_i^1 = n_{il} = \left. \bar{n}_i^0 \right|_{\bar{x}=-1+\epsilon \delta} \label{Eq: tilde-ni1-tD}. 
\end{eqnarray}
\end{subequations}
Eq. (\ref{Eq: tilde-ni0-tD}) is similar to the  result at the $t_{\lambda}$ scale. Therefore, simplifying Eq. (\ref{Eq: tilde-ni-tD}) at $O(1)$ yields that the Boltzmann distribution is valid, i.e., 
\begin{equation}
    \tilde{c}_i = \left. \bar{c}_i^0 \right|_{\bar{x}=-1+\epsilon \delta} \exp \left(-z_i \left( \tilde{\phi}^0 - \left. \bar{\phi}^0 \right|_{\bar{x}=-1+\epsilon \delta} \right) \right),
    \label{Eq: tilde-PB-tD} 
\end{equation}
where $\left. \bar{c}_i^0 \right|_{\bar{x}=-1+ \epsilon \delta}$ and  $\left. \bar{\phi}^0 \right|_{\bar{x}=-1 + \epsilon \delta}$ are to be determined from the bulk solution. We note that at the $t_{\lambda}$ scale, the concentration pre-factor in the Boltzmann distribution is the initial value. However, at the $t_D$ scale, this pre-factor is modified based upon changes in the reference point. Next, we focus on Eq. (\ref{Eq: tilde-ni-tD}), which states that the first-order flux is constant at the $t_D$ scale. Physically, since the electrical double layer is charged, there is no accumulation within the double layer, and thus the redox flux applied at the Stern-diffuse layer interface is also the flux at the diffuse layer-bulk interface. In other words, Eq. (\ref{Eq: tilde-ni1-tD}) provides a boundary condition at $\bar{x}=-1+ \epsilon \delta$ for Eq. (\ref{Eq: bar-ci0-tD}). 
\par{} Eq. (\ref{Eq: tilde-poisson-tD}) is identical to Eq. (\ref{Eq: tilde-poisson}). Following a similar process, it is straightforward to show
\begin{equation}
     \frac{\partial \tilde{\phi}^0}{\partial \tilde{x}} = \sqrt{2 \sum_i \left. \bar{c}_{i}^0 \right|_{\bar{x}=-1+\epsilon \delta} \left( \exp \left[ -z_i \left(\tilde{\phi}^0 - \left. \bar{\phi}^0 \right|_{\bar{x}=-1+\epsilon \delta} \right) \right] - 1  \right) }. 
     \label{Eq: dphitildedx-tD}
\end{equation}
Eq. (\ref{Eq: dphitildedx-tD}) enables us to use Eq. (\ref{Eq: tilde-stern-bc}) with the following definition of $\tilde{q}$ 
\begin{equation}
    \label{Eq: tilde-q-tD}
    \tilde{q} = \frac{\partial \tilde{\phi}^0}{\partial \tilde{x}}\bigg|_{\tilde{x}=0} = \sqrt{2 \sum_i \left. \bar{c}_{i}^0 \right|_{\bar{x}=-1+\epsilon \delta} \left( \exp \left[ -z_i \left(\tilde{\phi}^0(0,t_D) - \left. \bar{\phi}^0 \right|_{\bar{x}=-1+\epsilon \delta} \right) \right] - 1  \right) }.
\end{equation}
\par{} Finally, similar to Eq. (\ref{Eq: tilde-ci-tD}), perturbation analysis of Eq. (\ref{Eq: tilde-current-tD}) at $O(1)$ and $O(\epsilon)$ reveals $\tilde{j}^0=0$ and \begin{equation}
    \tilde{j}^1 = \left. \bar{j}^0 \right|_{\bar{x}=-1+\epsilon \delta} = j_l,  
    \label{Eq: tilde-jl1-tD}
\end{equation}
respectively. Therefore, Eq. (\ref{Eq: tilde-jl1-tD}) can serve as the boundary condition for Eq. (\ref{Eq: bar-j0-tD}). 
\subsection{Right electrical double-layer region}
The governing equations for the right electrical double-layer region become
\begin{subequations}
\begin{eqnarray}
\epsilon^2 \frac{\partial \hat{c}_i}{\partial t_D} +  \frac{\partial \hat{n}_i}{\partial \hat{x}} = 0, \label{Eq: hat-ci-tD}\\
\hat{n}_i = -\frac{\partial \hat{c}_i}{\partial \hat{x}} - z_i \hat{c}_i \frac{\partial \hat{\phi}}{\partial \hat{x}} \label{Eq: hat-ni-tD} \\
- \frac{\partial^2 \hat{\phi}}{\partial \hat{x}^2} = \hat{\rho}_e,  \label{Eq: hat-poisson-tD} \\
\epsilon^2 \frac{\partial \hat{\rho}_e}{\partial t_D} +  \frac{\partial \hat{j}}{\partial \hat{x}} = 0 \label{Eq: hat-current-tD}. 
\end{eqnarray}
\end{subequations}
Due to similarity with the left electrical double-layer region, we directly write the expressions here
\begin{subequations}
\begin{eqnarray}
 \hat{n}_i^0 = 0 \label{Eq: hat-ni0-tD}, \\
 \hat{n}_i^1 = n_{ir} = \left. \bar{n}_i^0 \right|_{\bar{x}=1-\epsilon \delta}, \label{Eq: hat-ni1-tD} 
\end{eqnarray}
\end{subequations}
\begin{equation}
    \hat{c}_i = \left. \bar{c}_i^0 \right|_{\bar{x}=1-\epsilon \delta} \exp \left(-z_i \left( \hat{\phi}^0 - \left. \bar{\phi}^0 \right|_{\bar{x}=1-\epsilon \delta} \right) \right),
    \label{Eq: hat-PB-tD} 
\end{equation}
\begin{equation}
     \frac{\partial \hat{\phi}^0}{\partial \hat{x}} = - \sqrt{2 \sum_i \left. \bar{c}_{i}^0 \right|_{\bar{x}=1-\epsilon \delta} \left( \exp \left[ -z_i \left(\hat{\phi}^0 - \left. \bar{\phi}^0 \right|_{\bar{x}=1-\epsilon \delta} \right) \right] - 1  \right) }, 
     \label{Eq: dphihatdx-tD}
\end{equation}
\begin{equation}
    \label{Eq: hat-q-tD}
    \hat{q} = \frac{\partial \hat{\phi}^0}{\partial \hat{x}}\bigg|_{\hat{x}=0} = -\sqrt{2 \sum_i \left. \bar{c}_{i}^0 \right|_{\bar{x}=1-\epsilon \delta} \left( \exp \left[ -z_i \left(\hat{\phi}^0(0,t_D) - \left. \bar{\phi}^0 \right|_{\bar{x}=1-\epsilon \delta} \right) \right] - 1  \right) }.
\end{equation}
\begin{equation}
    -\hat{j}^1 = \left. \bar{j}^0 \right|_{\bar{x}=1-\epsilon \delta} = j_r.   
    \label{Eq: hat-jl1-tD}
\end{equation}
Eqs. (\ref{Eq: bar-j0-tD}), (\ref{Eq: tilde-jl1-tD}), and (\ref{Eq: hat-jl1-tD}) demonstrate that the current is constant throughout the three spatial regions at each point in time. Therefore, at the $t_D$ scale, the net charge inside the whole system is zero, i.e., $\tilde{q} + \hat{q} = 0$, or $\left. \frac{\partial \tilde{\phi}^0}{\partial \tilde{x}} \right|_{\tilde{x}=0} + \left. \frac{ \partial \hat{\phi}^0}{\partial \hat{x}}\right|_{\hat{x}=0} = 0$. This condition, along with Stern layer boundary conditions similar to Eqs. (\ref{Eq: tilde-stern-bc}) and (\ref{Eq: hat-stern-bc}), enables us to write 
\begin{equation}
    \tilde{\phi}^0(0,t_D) + \hat{\phi}^0(0,t_D) = 0. 
    \label{Eq: tilde-hat-phi-sum}
\end{equation}

\subsection{Combining the three regions}
We solve for $\bar{c}_i^0(\bar{x},t_D)$, $\bar{\phi}^0(\bar{x},t_D)$, $\tilde{\phi}^0(0,t_D)$, $\hat{\phi}^0(0,t_D)$, $\tilde{q}(t_D)$, and $\hat{q}(t_D)$ by solving a system of partial differential equations and algebraic equations. Specifically, we use Eqs. (\ref{Eq: bar-ci0-tD}), (\ref{Eq: bar-ni0-tD}), (\ref{Eq: bar-j0-tD}), initial condition $\bar{c_{i}}^0(\bar{x},0)=c_{i0}$, and boundary/algebraic conditions provided in Eqs. (\ref{Eq: tilde-ni1-tD}), (\ref{Eq: hat-ni1-tD}), (\ref{Eq: tilde-jl1-tD}), (\ref{Eq: tilde-stern-bc}), (\ref{Eq: hat-stern-bc}), (\ref{Eq: tilde-q-tD}), (\ref{Eq: hat-q-tD}), and (\ref{Eq: tilde-hat-phi-sum}). Once $\bar{c}_i^0(\bar{x},t_D)$, $\bar{\phi}^0(\bar{x},t_D)$, $\tilde{\phi}^0(0,t_D)$, and $\hat{\phi}^0(0,t_D)$ are determined, Eqs. (\ref{Eq: dphitildedx-tD}) and (\ref{Eq: dphihatdx-tD}) can be numerically integrated with the known values of $\tilde{\phi}^0(0,t_D)$ and $\hat{\phi}^0(0,t_D)$. Finally, Eqs. (\ref{Eq: tilde-PB-tD}) and (\ref{Eq: hat-PB-tD}) can be used to find $\tilde{c}_i^0(\tilde{x},t_D)$ and $\hat{c}_i^0(\hat{x},t_D)$.   
\par{} Therefore, the spatially matched solution becomes
\begin{subequations}
\begin{equation}
\begin{aligned}
c_i^{D} (x, t_D) = \bar{c}_{i}^0(\bar{x},t_D) &+ \left. \bar{c}_i^0 \right|_{\bar{x}=-1+\epsilon \delta} \left( \exp \left[- z_i \left\{  \tilde{\phi}^0 \left( \frac{x + 1 - \epsilon \delta }{\epsilon}, t_D  \right) - \left. \bar{\phi}^0 \right|_{\bar{x}=-1 +\epsilon \delta} \right\}  \right]  -1 \right)  \\ 
    &+ \left. \bar{c}_{i}^0 \right|_{\bar{x}=1-\epsilon \delta}  \left( \exp \left[- z_i \left\{  \hat{\phi}^0 \left( \frac{x - 1 + \epsilon \delta}{\epsilon}, t_D \right) - \left. \bar{\phi}^0 \right|_{\bar{x}=1-\epsilon \delta} \right \} \right] -1 \right), 
\end{aligned}  
\end{equation}
\begin{equation}
    \phi^{D}(x,t_D) = \bar{\phi}^0 (\bar{x}, t_D) +  \left( \tilde{\phi}^0 \left( \frac{x + 1 - \epsilon \delta }{\epsilon}, t_D  \right) - \left. \bar{\phi}^0 \right|_{\bar{x}=-1+\epsilon \delta} \right) + \left( \hat{\phi}^0 \left( \frac{x - 1 + \epsilon \delta}{\epsilon}, t_D \right) - \left. \bar{\phi}^0 \right|_{\bar{x}=1-\epsilon \delta} \right).
\end{equation}
\label{Eq: D-full-sol}
\end{subequations}

\begin{table}[h!]
\caption{\textbf{Final equations for an arbitrary number of ions and redox reactions.} Summary of requisite equations to determine concentrations $c_i (x,t)$ and potential $\phi(x,t)$ for ion valences $z_i$ and reaction rates $n_{il}$ and $n_{ir}$ for an applied potential difference $2 \phi_s$. The ratio of Stern layer thickness $\lambda_s$ and diffuse layer length $\lambda$ is $\delta=\frac{\lambda_s}{\lambda}$. The equations are only valid when the separation between electrodes $L$ follows $\epsilon = \frac{\lambda}{L} \ll 1$. Further, the fluxes specified at the boundaries should not exceed the limiting current.}
\resizebox{0.95\textwidth}{!}{
\begin{tabular}{|c|cc|}
\hline & \multicolumn{1}{c|}{\textbf{{Short timescale}}}   & \textbf{{Long timescale}}  \\ \hline
\begin{tabular}[c]{@{}c@{}} \\ Step-1 \end{tabular} & \multicolumn{1}{c|}{\begin{tabular}[c]{@{}c@{}}\\ Solve for $A(t_\lambda)$, $B(t_\lambda)$, $\tilde{q}(t_\lambda)$, $\hat{q}(t_\lambda)$,  $\tilde{\gamma}(t_\lambda)$, and $\hat{\gamma}(t_\lambda)$ \\ $\displaystyle \delta \tilde{q} = \phi_s + \tilde{\gamma} - A + B$\\ \\ $\displaystyle \frac{d \tilde{q}}{d t_{\lambda}} = \sum_i z_i^2 c_{i0} A + j_l$\\ \\ $\displaystyle - \delta \hat{q} = \phi_s - \hat{\gamma} - A - B$\\ \\ $\displaystyle \frac{d \hat{q}}{d t_{\lambda}} = - \sum_i z_i^2 c_{i0} A - j_r$ \\ \\ $\tilde{q} = \sqrt{2 \sum_i c_{i0} \left( \exp \left[ -z_i \left(\tilde{\phi}^0 (0) + A - B \right) \right] - 1 \right) }$ \\ \\$\hat{q} = - \sqrt{2 \sum_i c_{i0} \left( \exp \left[ -z_i \left(\hat{\phi}^0 (0) - A - B \right) \right] - 1 \right) }$ \\ \\ with initial conditions \\ $A(0) = \phi_s$, $B(0) = 0$, \\ $\tilde{\gamma}(0) = 0$, $\hat{\gamma}(0) = 0$ \\ $\tilde{q}(0) = 0$, $\hat{q}(0) = 0$.\\ \end{tabular}} & \begin{tabular}[c]{@{}c@{}} Solve for $\bar{c_i}^0 (\bar{x},t_D)$, $\tilde{q}(t_D)$, $\hat{q}(t_D)$, $\bar{\phi}^0 (\bar{x},t_D)$, $\tilde{\phi}^0(0,t_D)$, and $\hat{\phi}^0(0,t_D)$ \\ $\displaystyle \frac{\partial \bar{c}_i^0}{\partial t_D} +  \frac{\partial \bar{n}_i^0}{\partial \bar{x}} = 0$; $\sum z_i \bar{c}^0_i=0$\\ \\ $\displaystyle \bar{n}_i^0 = - \frac{\partial \bar{c}_i^0}{\partial \bar{x}} - z_i \bar{c}_i^0 \frac{\partial \bar{\phi}^0}{\partial \bar{x}}$\\  \\  $\displaystyle  j_l = j_r = -\sum_i z_{i}^2 c_{i0} \frac{\partial \tilde{\phi}}{\partial \tilde{x}}$ \\ \\ $\tilde{q} = \sqrt{2 \sum_i \left. \bar{c}_{i}^0 \right|_{\bar{x}=-1+\epsilon \delta} \left( \exp \left[ -z_i \left(\tilde{\phi}^0(0,t_D) - \left. \bar{\phi}^0 \right|_{\bar{x}=-1+\epsilon \delta} \right) \right] - 1  \right) }$ \\ \\ $ \hat{q} = -\sqrt{2 \sum_i \left. \bar{c}_{i}^0 \right|_{\bar{x}=1-\epsilon \delta} \left( \exp \left[ -z_i \left(\hat{\phi}^0(0,t_D) - \left. \bar{\phi}^0 \right|_{\bar{x}=1-\epsilon \delta} \right) \right] - 1  \right) }$ \\  with initial conditions \\ 
$ \displaystyle \bar{c}_i^0 (\bar{x},0) = c_{i0}$  \\ 
and boundary/algebraic conditions  \\ $\displaystyle \tilde{\phi}^0(0,t_D) + \hat{\phi}^0(0,t_D) = 0$ \\ \\ $\displaystyle \left. \bar{n}_i^0 \right|_{\bar{x}=-1+\epsilon \delta} = n_{il} $\\ \\ $\displaystyle    \left. \bar{n}_i^0 \right|_{\bar{x}=1-\epsilon \delta} = n_{ir} $ \\ \\ $\delta \tilde{q} = \phi_s + \tilde{\phi}(0,t_D)$ \\ \\ $- \delta \hat{q} = \phi_s - \hat{\phi}(0,t_D)$ \\ \end{tabular} \\ \hline
\begin{tabular}[c]{@{}c@{}} Step-2 \end{tabular} & \multicolumn{1}{c|}{\begin{tabular}[c]{@{}c@{}}\\ Integrate equations \\ 
$\displaystyle \frac{\partial \tilde{\phi}^0}{\partial \tilde{x}} = \sqrt{2 \sum_i c_{i0} \left( \exp \left[ -z_i \left(\tilde{\phi}^0 + A - B \right) \right] - 1 \right) }$\\ \\ $\displaystyle \frac{\partial \hat{\phi}^0}{\partial \hat{x}} = - \sqrt{2 \sum_i c_{i0} \left( \exp \left[ -z_i \left(\hat{\phi}^0 - A - B \right) \right] - 1 \right) }$ \\ \\ with boundary conditions \\ $ \tilde{\phi}^0(0,t_\lambda) = \tilde{\gamma}(t_{\lambda}) - A(t_{\lambda}) + B(t_{\lambda})$, \\ \\ $\hat{\phi}^0(0,t_{\lambda}) = \hat{\gamma}(t_{\lambda}) + A(t_{\lambda}) + B(t_{\lambda})$. \end{tabular}}    & \begin{tabular}[c]{@{}c@{}} \\ Integrate equations \\ $\displaystyle \frac{\partial \tilde{\phi}^0}{\partial \tilde{x}} = \sqrt{2 \sum_i \left. \bar{c}_{i}^0 \right|_{\bar{x}=-1+\epsilon \delta} \left( \exp \left[ -z_i \left(\tilde{\phi}^0 - \left. \bar{\phi}^0 \right|_{\bar{x}=-1+\epsilon \delta} \right) \right] - 1  \right) }$\\ \\ $\displaystyle \frac{\partial \hat{\phi}^0}{\partial \hat{x}} = - \sqrt{2 \sum_i \left. \bar{c}_{i}^0 \right|_{\bar{x}=1-\epsilon \delta} \left( \exp \left[ -z_i \left(\hat{\phi}^0 - \left. \bar{\phi}^0 \right|_{\bar{x}=1-\epsilon \delta} \right) \right] - 1  \right) }$ \\ \\ with boundary conditions \\ $\displaystyle \left. \tilde{\phi}^0  \right|_{\tilde{x}=0} = \tilde{\phi}(0,t_D)$ from Step-1 \\ \\ $\displaystyle  \left. \hat{\phi}^0  \right|_{\hat{x}=0} = \hat{\phi}(0,t_D)$ from Step-1 \\\\\end{tabular}  \\ \hline
\begin{tabular}[c]{@{}c@{}} Space \\ stitching \end{tabular} & \multicolumn{1}{c|}{\begin{tabular}[c]{@{}c@{}} see Eq. (\ref{Eq: lamda-full-sol}) to obtain  $c_{i}^{\lambda}(x,t_{\lambda})$ and $\phi^{\lambda}(x,t_{\lambda})$\\\end{tabular}}    & \begin{tabular}[c]{@{}c@{}} see Eq. (\ref{Eq: D-full-sol}) to obtain $c_{i}^{D}(x,t_{D})$ and $\phi^{D}(x,t_D)$\\ \end{tabular}  \\ \hline
\begin{tabular}[c]{@{}c@{}}Time \\ stitching \end{tabular}& \multicolumn{2}{c|}{\begin{tabular}[c]{@{}c@{}} \\ $\displaystyle c_i (x, t) = c_i^{\lambda} \left(x, \frac{t}{\epsilon} \right) + c_i^D (x, t) - c_i^D (x, 0)$\\ \\ $\displaystyle \phi (x, t) = \phi^{\lambda} \left(x, \frac{t}{\epsilon} \right) + \phi^D (x, t) - \phi^D(x,0)$ \\ \\\end{tabular}}  \\ \hline
\end{tabular}
}
\label{tab:general}
\end{table}

\section{Combining short and long timescale solutions}
\label{sec: stitch}
Once the short and long timescale solutions are stitched spatially, it is possible to also combine them in time to get the final solution as follows:
\begin{subequations}
\begin{eqnarray}
    c_i (x, t) = c_i^{\lambda} \left(x, \frac{t}{\epsilon} \right) + c_i^D (x, t) - c_i^D (x, 0)  \\
    \phi (x, t) = \phi^{\lambda} \left(x, \frac{t}{\epsilon} \right) + \phi^D (x, t) - \phi^D(x,0). 
\end{eqnarray}
\label{Eq: full-soln}
\end{subequations}
Eq. (\ref{Eq: full-soln}) enables us to simulate coupled dynamics of electrical double layers and redox reactions for an arbitrary number of ions and redox reactions. The final equations are also summarized in Table \ref{tab:general}.

\section{Binary electrolyte with constant flux}
\label{sec: 2ion}
 We consider the system described in Fig. \ref{Fig: schematic} with a binary electrolyte, i.e., one cation and one anion.  We use subscript $i=+, -$ to denote properties of the cation and the anion, respectively. We assume $z_{\pm}= \pm 1$ and $c_{\pm0}=1$. We also assume that the cation is electrochemically active such $n_{+l}=n_{+r}=n$, where $n$ is a constant. We emphasize that the values of $n_l$ and $n_r$ could be dependent on potential and concentration, such as the relationships given by Frumkin-Butler-Volmer kinetics, but are assumed to be constant here for simplicity. Further, we assume that the anion does not undergo any redox reactions, i.e., $n_{-l} = n_{-r} = 0$. As a result, the Faradaic flux, i.e., $j_l = j_r = j_0$, where $j_0=n$ is constant; see Eqs. (\ref{Eq: tilde-j1-zero}) and (\ref{Eq: hat-j1-zero}). We note that the value of $j_0$ is arbitrary so long as it does not exceed the limiting current value, i.e., $|j_0| \le 2$. Physically, we also note that $j_0 <0$, i.e., the cation is being produced at the positive electrode and being consumed at the negative electrode. While the formulation would yield results for $j_0 > 0$, they are likely to be unphysical. This discrepancy gets automatically resolved when appropriate voltage-dependent fluxes are specified. 
\subsection{Short timescale solution}
\label{Sec: symm-short}
For the parameters stated above, Eqs. (\ref{Eq: tilde-final-Eqs}), (\ref{Eq: hat-final-Eqs}), (\ref{Eq: tilde-rhoe-dx}) and (\ref{Eq: hat-rhoe-dx}) reduce to 
\begin{subequations}
\label{Eq: tilde-hat-binary-all}
\begin{align}
- 2 \sqrt{2} \delta \sinh{\left(\frac{\tilde{\gamma}}{2}\right)} &= \phi_s + \tilde{\gamma} - A + B,
    \label{Eq: tilde-binary-Sternf} \\
\sqrt{2} \cosh{\left(\frac{\tilde{\gamma}}{2}\right)} \frac{d \tilde{\gamma}}{d t_{\lambda}} &= -2 A - j_0, \label{Eq: tilde-binary-Poissonf} \\
2 \sqrt{2} \delta \sinh{\left(\frac{\hat{\gamma}}{2}\right)} &= \phi_s - \hat{\gamma} - A - B, 
\label{Eq: hat-binary-Sternf}\\ 
\sqrt{2} \cosh{\left(\frac{\hat{\gamma}}{2}\right)} \frac{d \hat{\gamma}}{d t_{\lambda}} &= 2A + j_0,
\label{Eq: hat-binary-Poissonf} 
\end{align}
\end{subequations}
\noindent where $\tilde{\gamma} = \tilde{\phi}^0 (0,t_\lambda) + A - B$, $\hat{\gamma} = \hat{\phi}^0 (0,t_\lambda) - A - B$, $\tilde{q}=-2 \sqrt{2} \sinh{ \frac{\tilde{\gamma}}{2}}$, and $\hat{q}=-2 \sqrt{2} \sinh{ \frac{\hat{\gamma}}{2}}$. Adding Eqs. (\ref{Eq: tilde-binary-Poissonf}) and (\ref{Eq: hat-binary-Poissonf}), integrating, and applying the initial condition $\tilde{\gamma} (0) = \hat{\gamma} (0) = 0$ reveals
\begin{equation}
    \tilde{\gamma}(t_{\lambda}) + \hat{\gamma}(t_{\lambda}) = 0.
    \label{Eq: gammahat+gammatilde}
\end{equation}
Subtracting Eq. (\ref{Eq: hat-binary-Sternf}) from Eq. (\ref{Eq: tilde-binary-Sternf}) and substituting the result given from Eq. (\ref{Eq: gammahat+gammatilde}) yields
\begin{equation}
    B (t_\lambda) = 0.
\end{equation} 
Thus, Eqs. (\ref{Eq: tilde-hat-binary-all}) simplify to 
\begin{subequations}
\label{Eq: binary-all}
\begin{align}
    - 2 \sqrt{2} \delta \sinh{\left(\frac{\tilde{\gamma}}{2}\right)} &= \phi_s + \tilde{\gamma} - A, \\
    \sqrt{2} \cosh{\left(\frac{\tilde{\gamma}}{2}\right)} \frac{d \tilde{\gamma}}{d t_{\lambda}} &= -2A - j_0,
\end{align}
which are solved numerically using the \textit{ode15s} utility in MATLAB with $\tilde{\gamma}(0)=0$ and $A(0)=\phi_s$ to obtain $\tilde{\gamma}(t_{\lambda})$ and $A(t_{\lambda})$. $\hat{\gamma}(t_{\lambda})$ can be obtained by using Eq. (\ref{Eq: gammahat+gammatilde}). Finally, to obtain concentration and potential profiles inside the double layers, Eq. (\ref{Eq: dphitildedx}) and (\ref{Eq: dphihatdx}) are integrated to get
\begin{align}
     \tanh{\frac{\tilde{\phi}^0 + A}{4}} = \tanh{\frac{\tilde{\gamma}}{4}} \exp(-\sqrt{2} \tilde{x}),
     \label{Eq: dphitildedxfinal} \\
          \tanh{\frac{\hat{\phi}^0 - A}{4}} = \tanh{\frac{\hat{\gamma}}{4}} \exp(-\sqrt{2} \hat{x}).
     \label{Eq: dphihatdxfinal}
\end{align}
\end{subequations}
Eqs. (\ref{Eq: binary-all}), combined with Eq. (\ref{Eq: gammahat+gammatilde}), enable us to evaluate all the terms described in Eq. (\ref{Eq: lamda-full-sol}) and thus determine the complete solution at the short timescale, i.e., $c_{\pm}^{\lambda} (x, t_{\lambda})$ and $\phi^{\lambda}(x,t_\lambda)$.

\subsection{Long timescale solution}
\label{Sec: symm-long}
For the parameters stated above, Eq. (\ref{Eq: bar-poisson-tD}) at $O(1)$ reveals $\bar{\rho}_e^0= 0$. Therefore, $\bar{c}_{+}^0 = \bar{c}_{-}^0 = \bar{c}^0$. Writing Eqs. (\ref{Eq: bar-ci0-tD}) and (\ref{Eq: bar-ni0-tD}) for both ions and adding them yields
\begin{equation}
    \frac{\partial \bar{c}^0}{\partial t_D}= \frac{\partial^2 \bar{c}^0}{\partial \bar{x}^2},
    \label{Eq: bar-bulk-diffusion}
\end{equation}
where $\bar{c}^0(\bar{x},0) = 1$, i.e., the steady state solution at the $t_{\lambda}$ scale becomes the initial condition at the $t_D$ scale. Boundary conditions for Eq. (\ref{Eq: bar-bulk-diffusion}) are determined by utilizing Eqs. (\ref{Eq: tilde-ni1-tD}) and (\ref{Eq: hat-ni1-tD}), which for each ion yields $\left. \frac{\partial \bar{c}^0}{\partial \bar{x}} \right|_{\bar{x}=\pm 1 \mp \epsilon \delta} = -\frac{j_0}{2}$. Eq. (\ref{Eq: bar-bulk-diffusion}) has  a series solution \cite{deen2012analysis} 
\begin{equation}
    \bar{c}^0 (\bar{x}, t_D) = 1 - \frac{j_0 \bar{x}}{2} - 2 j_0 \sum_{n=1} ^{\infty}{ \frac{ (1-(-1)^n) }{n^2 \pi^2} \cos{\frac{n \pi (\bar{x} + 1)}{2}} \exp\left( - \frac{n^2 \pi^2}{4} t_D\right)}.
    \label{Eq: FFT-solution}
\end{equation}
Eq. (\ref{Eq: FFT-solution}) provides the solution for $\bar{c}^0 (\bar{x},t_D)$. To determine $\bar{\phi}^0(\bar{x},t_D)$, we utilize Eq. (\ref{Eq: bar-j0-tD}) to write
\begin{equation}
    \bar{c}^0 \frac{\partial \bar{\phi}^0}{\partial \bar{x}}= -\frac{j_0}{2}.
    \label{Eq: bar-electromigration}
\end{equation}
In order to fully determine $\bar{\phi}^0(\bar{x},t_D)$, we need an additional boundary condition. To do so, we apply the boundary conditions given by Eqs. (\ref{Eq: tilde-stern-bc}), (\ref{Eq: hat-stern-bc}), and (\ref{Eq: tilde-hat-phi-sum}), which reduce to \begin{subequations}
\label{Eq: phi-bar-bc}
\begin{eqnarray}
    -2 \sqrt{2} \delta \sqrt{\left.\bar{c}^0 \right|_{\bar{x} = -1+\epsilon \delta}} \sinh{ \left(\frac{\left.\tilde{\phi}^0 \right|_{\tilde{x} = 0} - \left. \bar{\phi}^0 \right|_{\bar{x} = -1 + \epsilon \delta}}{2} \right)} &= \left.\tilde{\phi}^0 \right|_{\tilde{x} = 0} + \phi_s 
    \label{Eq: tilde-binary-Sternf-tD}  \\
     2 \sqrt{2} \delta \sqrt{\left.\bar{c}^0 \right|_{\bar{x} = 1-\epsilon \delta}} \sinh{ \left(\frac{\left.\hat{\phi}^0 \right|_{\hat{x} = 0} - \left. \bar{\phi}^0 \right|_{\bar{x} = 1-\epsilon \delta}}{2} \right)} &= \left.\hat{\phi}^0 \right|_{\hat{x} = 0} - \phi_s ,
    \label{Eq: hat-binary-Sternf-tD} \\
    \left. \tilde{\phi}^0 \right|_{\tilde{x}=0} +     \left. \hat{\phi}^0 \right|_{\hat{x}=0} = 0,
    \label{Eq: Fig4a}
\end{eqnarray}
\end{subequations}
where we have utilized the expressions for $\tilde{q}$ and $\hat{q}$ from Eqs. (\ref{Eq: tilde-q-tD}) and (\ref{Eq: hat-q-tD}). Eqs. (\ref{Eq: phi-bar-bc}) provide us a system of 3 equations with two additional unknowns $\tilde{\phi}(0,t_D)$ and $\hat{\phi}(0,t_D)$. Therefore, Eq. (\ref{Eq: bar-electromigration}) is integrated numerically by using the solution from Eq. (\ref{Eq: FFT-solution}) and employing Eqs. (\ref{Eq: phi-bar-bc}) as boundary conditions to obtain $\bar{\phi}^0(\bar{x},t_D)$, $\tilde{\phi}(0,t_D)$ and $\hat{\phi}(0,t_D)$. To numerically integrate, we discretize the equation in $\bar{x}$ and use the \textit{ode15s} utility in MATLAB. 

\begin{table}[h!]
\caption{\textbf{Final equations for a binary electrolyte with a constant redox flux}. The table below presents a summary of the equations that can be used to determine the concentrations $c_{\pm}(x,t)$ and potential $\phi(x,t)$ for a model system with only the cation reacting at a constant rate. Mathematically, we assume $c_{\pm}(x,0)=1$, $z_{\pm}= \pm 1$, $n_{+l}=n_{+r}=j_0$ and $n_{-l}=n_{-r}=0$. Note that the limiting current condition requires $|j_0| \le 2$. }
\label{tab:binary}
\resizebox{0.95\textwidth}{!}{
\begin{tabular}{|c|cc|}
\hline
 & \multicolumn{1}{c|}{ \textbf{Short timescale} }     &  \textbf{Long timescale}\\ \hline
\begin{tabular}[c]{@{}c@{}}Solution\\for\\individual\\timescales\end{tabular} & \multicolumn{1}{c|}{\begin{tabular}[c]{@{}c@{}}\\ Solve for $A(t_\lambda)$ and $\tilde{\gamma} (t_\lambda)$ \\[5pt] $\displaystyle \sqrt{2} \cosh{\left(\frac{\tilde{\gamma}}{2}\right)} \frac{d \tilde{\gamma}}{d t_{\lambda}} = -2A - j_0$ \\ $\displaystyle -2\sqrt{2} \delta \sinh{\left(\frac{\tilde{\gamma}}{2}\right)} = \phi_s + \tilde{\gamma} - A$\\ \\ with initial conditions \\ $A(0) = \phi_s$ \\ $\tilde{\gamma}(0) = 0$ \\ \\ and use relations\\ $\displaystyle \tilde{\gamma}(t_\lambda) + \hat{\gamma}(t_\lambda) = 0$ \\
$\displaystyle B(t_\lambda)= 0$. \\
\\ Next, solve for $\tilde{\phi}^0 (\tilde{x},t_{\lambda})$ and $\hat{\phi}^0 (\hat{x},t_{\lambda})$ \\ $\displaystyle \tanh{\frac{\tilde{\phi}^0 + A}{4}} = \tanh{\frac{\tilde{\gamma}}{4}} \exp(-\sqrt{2} \tilde{x})$\\ $\displaystyle \tanh{\frac{\hat{\phi}^0 - A}{4}} = \tanh{\frac{\hat{\gamma}}{4}} \exp(-\sqrt{2} \hat{x})$ \\ \\ \end{tabular}}                                 & \begin{tabular}[c]{@{}c@{}}\\ 
Use $\bar{c}^0 (\bar{x},t_D)$  \\ $\displaystyle \bar{c}^0 (\bar{x}, t_D) = 1 - \frac{j_0 \bar{x}}{2} -  2 j_0 S $\\
$\displaystyle S = \sum_{n=1} ^{\infty}{ \frac{ (1-(-1)^n) }{n^2 \pi^2} \cos{\frac{n \pi (\bar{x} + 1)}{2}} \exp\left( - \frac{n^2 \pi^2}{4} t_D\right)}$ \\
\\ Next, solve for $\bar{\phi}^0 (\bar{x},t_D)$, $\tilde{\phi}^0(0,t_D)$, and $\hat{\phi}^0(0,t_D)$ using \\ $\displaystyle \bar{c}^0 \frac{\partial \bar{\phi}^0}{\partial \bar{x}}= -\frac{j_0}{2}$\\ \\ with boundary/algebraic conditions \\ $\displaystyle -2\sqrt{2} \delta \sqrt{\left.\bar{c}^0 \right|_{\bar{x} = -1+\epsilon \delta}} \sinh{ \left(\frac{\left.\tilde{\phi}^0 \right|_{\tilde{x} = 0} - \left. \bar{\phi}^0 \right|_{\bar{x} = -1+\epsilon \delta}}{2} \right)} = \left.\tilde{\phi}^0 \right|_{\tilde{x} = 0} + \phi_s$\\ \\ $\displaystyle 2\sqrt{2} \delta \sqrt{\left.\bar{c}^0 \right|_{\bar{x} = 1-\epsilon \delta}} \sinh{ \left(\frac{\left.\hat{\phi}^0 \right|_{\hat{x} = 0} - \left. \bar{\phi}^0 \right|_{\bar{x} = 1-\epsilon \delta}}{2} \right)} = \left.\hat{\phi}^0 \right|_{\hat{x} = 0} - \phi_s$ \\ \\ $\displaystyle \left. \tilde{\phi}^0 \right|_{\tilde{x}=0} + \left. \hat{\phi}^0 \right|_{\hat{x}=0} = 0$. \\ \\ Next, solve for $\tilde{\phi}^0(\tilde{x},t_D)$ and $\hat{\phi}^0(\hat{x},t_D)$ using \\ $\displaystyle \tanh{\frac{\tilde{\phi}^0 - \left. \bar{\phi}^0 \right|_{\bar{x}=-1+\epsilon \delta}}{4}} = \tanh{\frac{\left(\left. \tilde{\phi^0} \right|_{\tilde{x} = 0} - \left. \bar{\phi}^0  \right|_{\bar{x}=-1+\epsilon \delta}\right)}{4}} \exp{\left( -\sqrt{2} \tilde{x} \sqrt{\left. \bar{c}^0 \right|_{\bar{x}=-1+\epsilon \delta}} \right)}$\\ $\displaystyle \tanh{\frac{ \hat{\phi}^0 - \left. \bar{\phi}^0 \right|_{\bar{x}=1-\epsilon \delta}}{4}} = \tanh{\frac{\left(\left. \hat{\phi}^0 \right|_{\hat{x}=0} - \left. \bar{\phi}^0  \right|_{\bar{x}=1-\epsilon \delta}\right)}{4}} \exp{\left( -\sqrt{2} \hat{x} \sqrt{ \bar{c}^0 |_{\bar{x}=1-\epsilon \delta}} \right)}$. \\ \\\end{tabular}\\ \hline
\multicolumn{1}{|l|}{\begin{tabular}[c]{@{}c@{}}Spatial\\ stitching\end{tabular}} & \multicolumn{1}{l|}{\begin{tabular}[c]{@{}c@{}} See Eq. (\ref{Eq: lamda-full-sol}) to obtain \\ $c_{\pm}^{\lambda}(x,t_{\lambda})$ and $\phi^{\lambda}(x,t_{\lambda})$ \end{tabular}} & \multicolumn{1}{l|}{\begin{tabular}[c]{@{}c@{}} See Eq. (\ref{Eq: D-full-sol}) to obtain $c_{\pm}^{D}(x,t_{D})$ and $\phi^{D}(x,t_{D})$ \end{tabular}} \\ \hline
\begin{tabular}[c]{@{}c@{}}Time\\stitching\end{tabular}                & \multicolumn{2}{c|}{\begin{tabular}[c]{@{}c@{}}\\$\displaystyle c_{\pm} (x, t) = c_{\pm}^{\lambda} \left(x, \frac{t}{\epsilon} \right) + c_{\pm}^D (x, t) - c_{\pm}^D (x, 0)$\\ \\ $\displaystyle \phi (x, t) = \phi^{\lambda} \left(x, \frac{t}{\epsilon} \right) + \phi^D (x, t) - \phi^D(x,0)$ \\ \\ \end{tabular}}\\ \hline
\end{tabular}
}
\end{table}

To obtain potential profiles within the double layers, we integrate Eqs. (\ref{Eq: dphitildedx-tD}) and (\ref{Eq: dphihatdx-tD}) to get 
\begin{subequations}
\begin{eqnarray}
\tanh{\frac{\tilde{\phi}^0(\tilde{x},t_D) - \left. \bar{\phi}^0 \right|_{\bar{x}=-1+\epsilon \delta}}{4}} = \tanh{\frac{\left(\left. \tilde{\phi^0} \right|_{\tilde{x} = 0} - \left. \bar{\phi}^0  \right|_{\bar{x}=-1+\epsilon \delta}\right)}{4}} \exp{\left( -\sqrt{2} \tilde{x} \sqrt{\left. \bar{c}^0 \right|_{\bar{x}=-1+\epsilon \delta}} \right)},
\label{Eq: dphitildedxfinall} \\
\tanh{\frac{ \hat{\phi}^0(\hat{x},t_D) - \left. \bar{\phi}^0 \right|_{\bar{x}=1-\epsilon \delta}}{4}} = \tanh{\frac{\left(\left. \hat{\phi}^0 \right|_{\hat{x}=0} - \left. \bar{\phi}^0  \right|_{\bar{x}=1-\epsilon \delta}\right)}{4}} \exp{\left( -\sqrt{2} \hat{x} \sqrt{ \bar{c}^0 |_{\bar{x}=1-\epsilon \delta}} \right)}.
\label{Eq: dphihatdxfinalr}
\end{eqnarray}
\label{Eq: phi-tilde-hat}
\end{subequations}
Eqs. (\ref{Eq: FFT-solution}), (\ref{Eq: bar-electromigration}), (\ref{Eq: phi-bar-bc}) and (\ref{Eq: phi-tilde-hat}) enable us to obtain the full solution at this timescale, i.e., $c_{\pm}^D (x, t_D)$ and $\phi^D(x,t_D)$, as described in Eq. (\ref{Eq: D-full-sol}). 
\par{} After obtaining $c_{\pm}^\lambda, \phi^\lambda, c_{\pm}^D$ and $\phi^D$, we combine the solutions at two timescales using Eq. (\ref{Eq: full-soln}). The final equations are summarized in Table \ref{tab:binary}. 

\subsection{Details of Direct Numerical Solution}
\label{Sec: symm-DNS}
The direct numerical solution of the PNP equations is challenging since the equations are coupled and highly non-linear, with strategies such as Quadtree adaptive grids having been proposed to solve these equations \cite{gibou2013high,mirzadeh2014enhanced}. Therefore, to solve the PNP equations numerically, we discretize the equations in space using a 6$^{\textrm{th}}$-order compact finite-difference scheme for a non-uniform grid \cite{lele1992compact,gamet1999compact}. We mesh the solution with a non-uniform grid such that there is sufficient resolution within the diffuse layer region. To perform integration in time, we employ the \textit{ode15s} utility in MATLAB. We benchmark our numerical code with the results provided for $j_0=0$ in Bazant et al. \cite{bazant2004diffuse}.
\subsection{Results and Discussion}
\begin{figure}[h!]
    \centering
    \includegraphics[width=\textwidth,keepaspectratio]{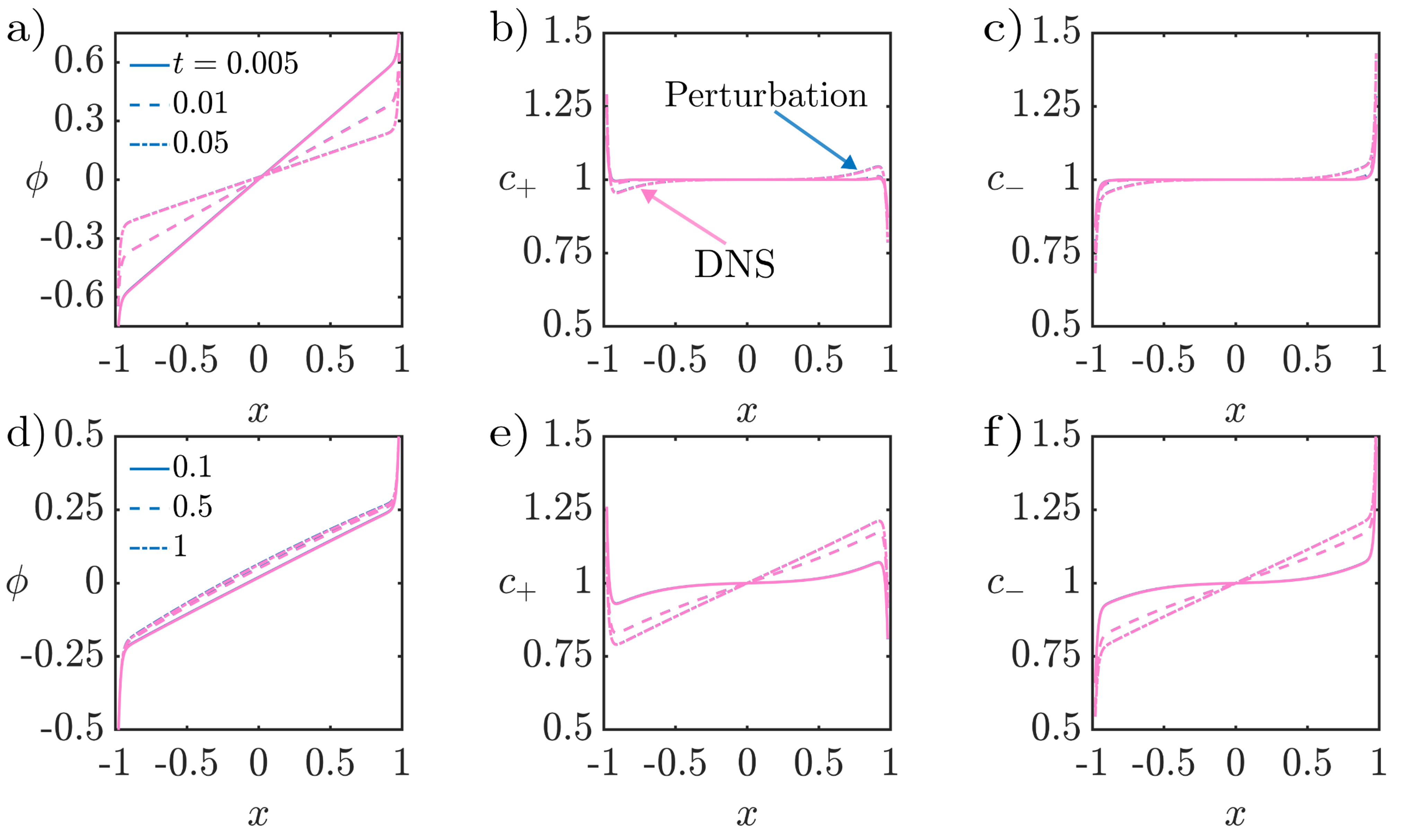}
    \caption{\textbf{Quantitative comparison between DNS and perturbation model.} DNS (pink) and perturbation results (blue) for potential $\phi$ and ion concentrations $c_\pm$. (a-c) $t = 0.005, 0.01, 0.05$ (d-f) $t = 0.1, 0.5, 1$. Excellent quantitative agreement is obtained across space and time. $\phi_s = 1$, $\delta = 1$, $\epsilon = 0.02$, and $j_0 = -0.5$.}
    \label{Fig: DNS}
\end{figure}

In this subsection, we compare the results for a symmetric electrolyte with constant flux from our proposed theoretical framework (see Sections \ref{Sec: symm-short} and \ref{Sec: symm-long}) with direct numerical simulations, abbreviated here as DNS (see Section \ref{Sec: symm-DNS}). Fig. \ref{Fig: DNS} displays a comparison between the two approaches for electrical potential $\phi(x, t)$ and concentration of ions $c_{\pm}(x, t)$; also see supplementary video 1. We note that the perturbation model (blue) displays excellent quantitative agreement for $\phi(x,t)$ and $c_{\pm}(x,t)$ with the DNS (pink) and is able to capture variations in space, including both the double layer and the bulk regions, at all times; see Fig. \ref{Fig: DNS}(a-f).

\par{} We first detail the results for $\phi(x,t)$. We note that Fig. \ref{Fig: DNS}(a) shows the results for short timescales, i.e., $t=0.005,0.01,0.05$, which is equivalent to $t_{\lambda}= \frac{t}{\epsilon}=0.25,0.5,2.5$. Therefore, the potential profiles in Fig. \ref{Fig: DNS}(a) correspond to the double-layer charging timescale. We find that at this timescale, the potential drop is linear in the bulk region, but the slope changes rapidly in time. Additionally, the potential drop across the double layers also changes with time, which is expected at the short timescale. In contrast, Fig. \ref{Fig: DNS}(d) shows results for the long timescale, i.e., $t=0.1,0.5,1$, which is representative of the bulk diffusion timescale. Here, $\phi(x,t)$ starts to deviate from the linear profile and develops a curvature in the bulk section. This curvature occurs because the potential gradient is required to allow the electromigrative and diffusive fluxes to together equal the surface reactive flux at long times while maintaining electroneutrality. We highlight that the potential drops across the left and right double layers are weakly dependent on time. 

\par{} Next, we focus on the predictions of $c_{\pm}(x,t)$. At the short timescale, i.e., Fig. \ref{Fig: DNS} (b-c), the ion concentrations in the bulk remain constant. In the double-layer regions, the positive ion concentration increases at the negative electrode whereas it decreases at the positive electrode. This trend is opposite for the negative ion. However, we note that at $t=0.05$, i.e., $t_{\lambda}=\frac{t}{\epsilon}=2.5$, there are small deviations from the constant value in the bulk that start to appear. These deviations occur because once the EDLs are fully charged, surface reactions start to impact the bulk region, leading to a finite accumulation of ions. We emphasize that this change does not occur in the absence of a surface reaction flux \cite{bazant2004diffuse, kilic2007steric, kilic2007stericpartone}, which highlights that the double-layer charging is dependent on the Faradaic process. In fact, this coupling becomes pronounced at the long timescale, i.e., 
Fig. \ref{Fig: DNS}(e-f). We find that the concentration profiles in the bulk region begin to curve at intermediate times and eventually become linear as the system approaches steady state, consistent with the steady state predictions of Bazant et al. \cite{bazant2005current}. Physically, concentrations start to change in the bulk as the effect of surface reactions moves beyond the double-layer regions, which leads to accumulation of ions in the bulk. Over time, the transport of ions through diffusion and electromigration adjust themselves so as to equal the surface Faradaic flux, and eventually reach a steady state. Further, we observe that the concentration of ions in the double layers also gets modified significantly at this timescale, which signifies a change in the double-layer structure, as we detail below. 

\begin{figure}[h!]
    \centering
    \includegraphics[width = \textwidth, keepaspectratio]{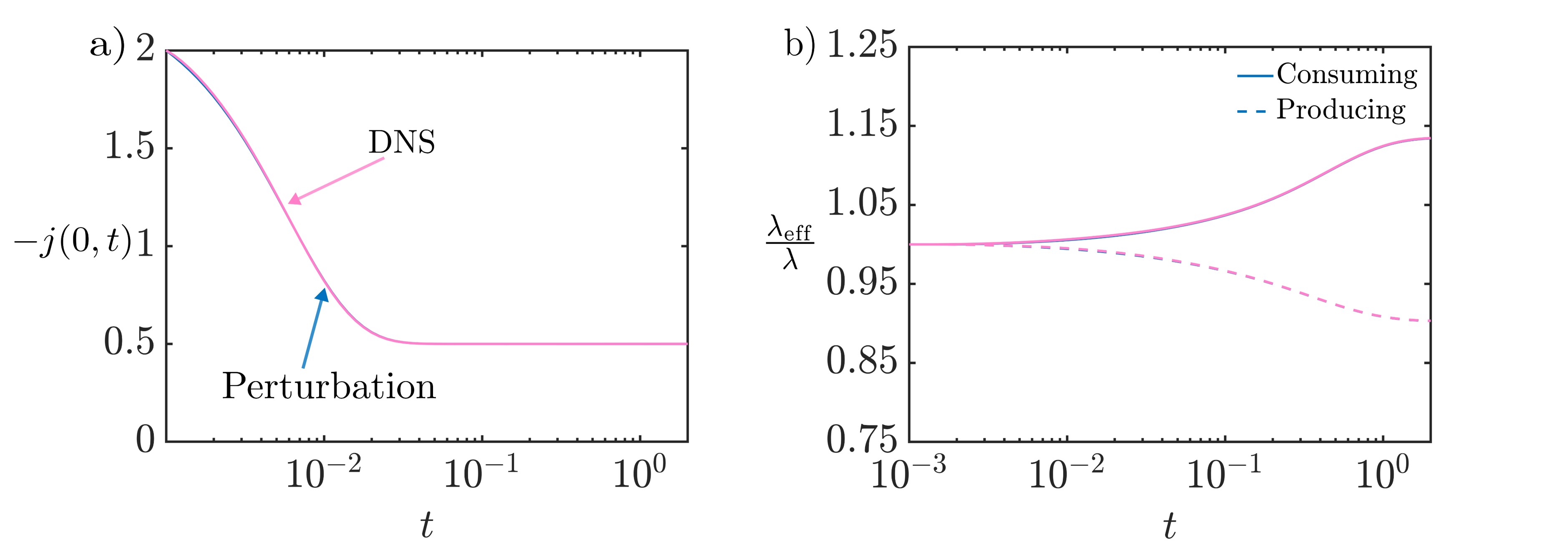}
    \caption{{\textbf{Dynamics of centerline current and effective Debye length.} (a) Magnitude of the centerline current $-j(0,t)$ vs. $t$ evaluated from DNS (pink) and perturbation model (blue). (b) Ratio of effective Debye length to initial Debye length $\frac{\lambda_{\mathrm{eff}}}{\lambda}$ vs. $t$ near reacting (solid) and producing electrodes (dashed) from DNS (pink) and perturbation model (blue). The results display quantitative agreement between the DNS and the proposed perturbation model. The centerline current decreases at short times and equals $j_0$ at long times. $\lambda_{\mathrm{eff}}$ decreases at the producing electrode due to an increase in local ion concentration and increases at the consuming electrode due to a decrease in local ion concentration.   $\phi_s = 1$, $\delta = 1$, $\epsilon = 0.02$, and $j_0 = -0.5$.}}
    \label{Fig: current}
\end{figure}

\par{}To dive deeper into the physics of the problem, we evaluate the centerline current $-j(0,t) = \left. 2 c \frac{\partial \phi}{\partial x}\right|_{x=0}$; see Fig. \ref{Fig: current}(a). Comparison between DNS and the proposed perturbation model displays quantitative agreement across all timescales. The centerline current is maximal at $t = 0$ and rapidly decreases at short times, and eventually relaxes to the constant $j_0$ set by the surface reactive flux. The time at which the centerline current is equal to $j_0$ is approximately $t = 0.03$, or $t_{\lambda}=\frac{t}{\epsilon}=1.5$, i.e., the time at which surface reactions start to impact the bulk. Our model is consistent with the predictions of Bazant et. al. \cite{bazant2004diffuse} for ideally blocking electrodes, i.e., if $j_0 = 0$, the current will vanish at the bulk diffusion timescale. The current at $t = 0$ is given by the short timescale solution, i.e., $j(0,0) = -2A(0) =-2\phi_s$. As the double layers get charged, this current rapidly decreases as $A(t_{\lambda})$ decreases. At the $t_D$ scale, however, our perturbation analysis (see Section \ref{sec: long}) shows that the current in each EDL is required to be equal to $j_0$, which means that the current in the bulk is also $j_0$. 

\par{}We now focus on Fig. \ref{Fig: current}(b), which displays changes with time of the effective double-layer thickness $\lambda_{\mathrm{eff}}$, at both consuming and producing electrodes, relative to the initial value of $\lambda$. To estimate $\frac{\lambda_{\mathrm{eff}}}{\lambda}$, we first find the local maximum of $\bar{c}_+$ near the producing electrode and the minimum of $\bar{c}_+$ near the consuming electrode, then employ $\frac{\lambda_{\mathrm{eff}}}{\lambda} = \frac{1}{\sqrt{\bar{c}_\mathrm{+,max/min}}}$ for each point in time. We find an increase in $\lambda_{\mathrm{eff}}$ near the consuming electrode and a decrease in $\lambda_{\mathrm{eff}}$ near the producing electrode. Importantly, $\lambda_{\mathrm{eff}}$ stays roughly constant at the short timescale, though small deviations start to appear at $t_{\lambda} \approx 0.5$. At the bulk diffusion timescale, there is a more pronounced shift in the value of $\lambda_{\mathrm{eff}}$. These changes in $\lambda_{\mathrm{eff}}$ have a direct dependence on $j_0$, which implies that control of electrochemical reaction rate could be used to control the size of EDLs. As a consequence of controlling $\lambda_{\mathrm{eff}}$, the  capacitance of the EDLs can also be influenced and can be different at the two electrodes.

\begin{figure}[b!]
    \centering
    \includegraphics[width = \textwidth, keepaspectratio]{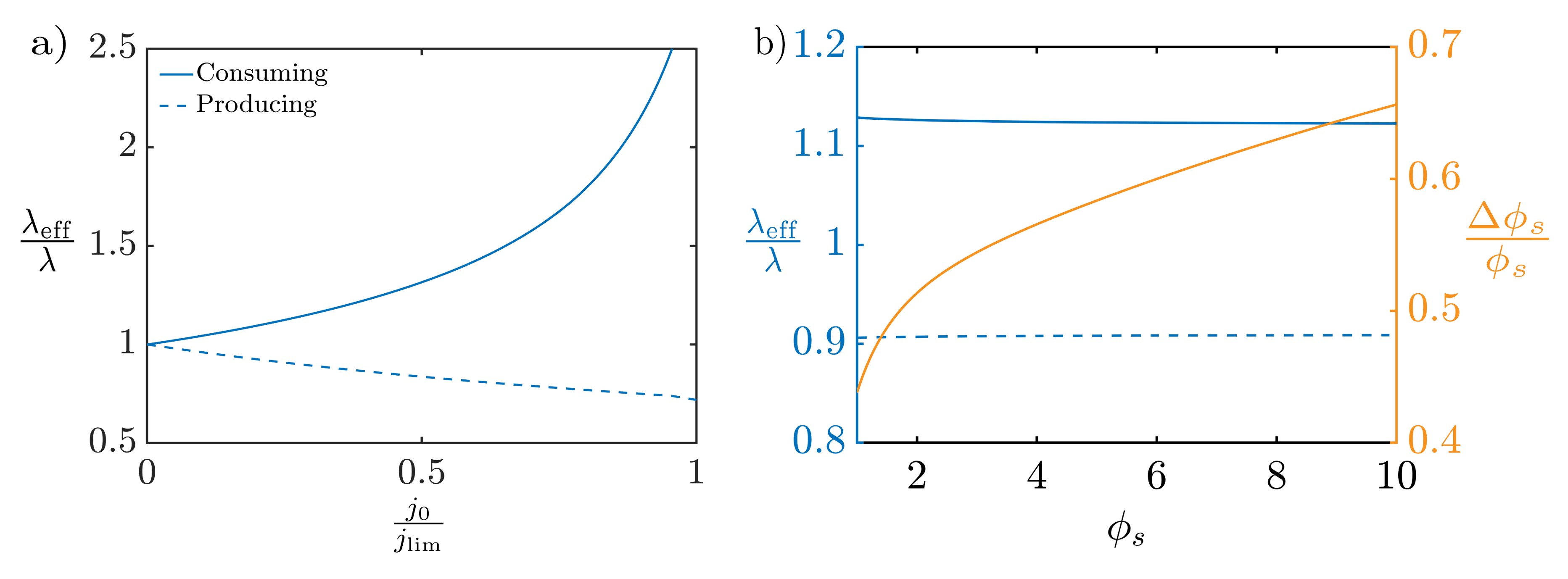}
    \caption{{\textbf{Dependence of effective Debye length on $\phi_s$ and $j_0$.} (a) $\frac{\lambda_{\mathrm{eff}}}{\lambda}$ vs. the ratio of $j_0$ to the limiting current $\frac{j_0}{j_{\mathrm{lim}}}$ with $\phi_s = 1$ and $t= 1$. (b) Left y-axis: $\frac{\lambda_{\mathrm{eff}}}{\lambda}$ vs. $\phi_s$ near consuming (blue, solid) and producing (blue, dashed) electrodes. Right y-axis: relative potential drop across the Stern layer $\frac{\Delta \phi_s}{\phi_s}$ vs. $\phi_s$ for consuming (orange, solid) and producing (orange, dashed) electrodes. The orange solid and dashed lines coincide, see Eq. (\ref{Eq: Fig4a}). $j_0 = -0.5$ and $t= 1$. $\frac{\lambda_{\mathrm{eff}}}{\lambda}$ is weakly dependent on $\phi_s$, and significantly dependent upon $j_0$. In both subfigures, $\delta = 1$ and $\epsilon = 0.02$.}}
    \label{Fig: lambdaratios}
\end{figure}

\par{} Next, we investigate the variation of $\lambda_{\mathrm{eff}}$ with $j_0$ at long times while keeping $\phi_s$ to be constant; see Fig. \ref{Fig: lambdaratios}(a). With an increase in $j_0$, $\lambda_{\mathrm{eff}}$ deviates further from unity at both electrodes. Quantitatively, we can approximate the value of $\lambda_{\mathrm{eff}} = \frac{1}{\sqrt{\bar{c}(\mp 1 \pm \epsilon \delta, t_D)}}$ for consuming and producing electrodes, respectively. At steady state, by using Eq. (\ref{Eq: FFT-solution}), we find $\frac{\lambda_{\mathrm{eff}}}{\lambda} = \frac{1}{\sqrt{1\mp \frac{j_0}{j_{\mathrm{lim}}}}}$. As $\frac{j_0}{j_{\mathrm{lim}}} \rightarrow 1$,  $\frac{\lambda_{\mathrm{eff}}}{\lambda} \rightarrow \frac{1}{\sqrt{2}}$ at the producing electrode and  $\frac{\lambda_{\mathrm{eff}}}{\lambda} \rightarrow \infty$ at the consuming electrode. In this limit, the EDLs extend into the bulk, which implies that the asymptotic ordering of the perturbation expansion breaks down and the model is no longer valid. However, our model remains valid for a wide range of currents, up to roughly 90-95 percent of the limiting current. 

\par{} We now study the dependence of $\lambda_{\mathrm{eff}}$ on $\phi_s$; see Fig \ref{Fig: lambdaratios}(b), left y-axis. Due to the assumption that $j_0$ is constant, unsurprisingly, we find that $\frac{\lambda_{\mathrm{eff}}}{\lambda}$ is weakly dependent upon $\phi_s$. In realistic systems, it is known that $j_0$ could be a function of $\phi_s$, which could induce a dependence of $\frac{\lambda_{\mathrm{eff}}}{\lambda}$ on $\phi_s$. We emphasize that our framework includes such dependences, but for simplicity, we do not investigate this effect in this section. At large $\phi_s$ values, another consideration is the depletion of ions from the bulk at the EDL charging timescale. Physically, for large potential drops across the diffuse layers, ions at \textit{both} electrodes deplete due to electrostatic interactions \cite{bazant2004diffuse}. Due to the inclusion of the Stern layer, at large $\phi_s$ values, the bulk of the potential drop occurs across the Stern layers $\Delta \phi_s$ seen in Fig. \ref{Fig: lambdaratios}(b), right y-axis. Therefore, our model remains valid for large applied potentials for sufficiently thick Stern layers. Interestingly, the potential drop across the Stern layers at the two electrodes are identical, driven by the requirement of equal and opposite charges inside the two EDLs; see Eqs. (\ref{Eq: hat-jl1-tD}) and (\ref{Eq: tilde-hat-phi-sum}). We note that the potential drops across the diffuse layers are not equal, leading to an asymmetry in potential profiles; see Fig. \ref{Fig: DNS}(d).   

\begin{figure}[b!]
    \centering
    \includegraphics[width=7in]{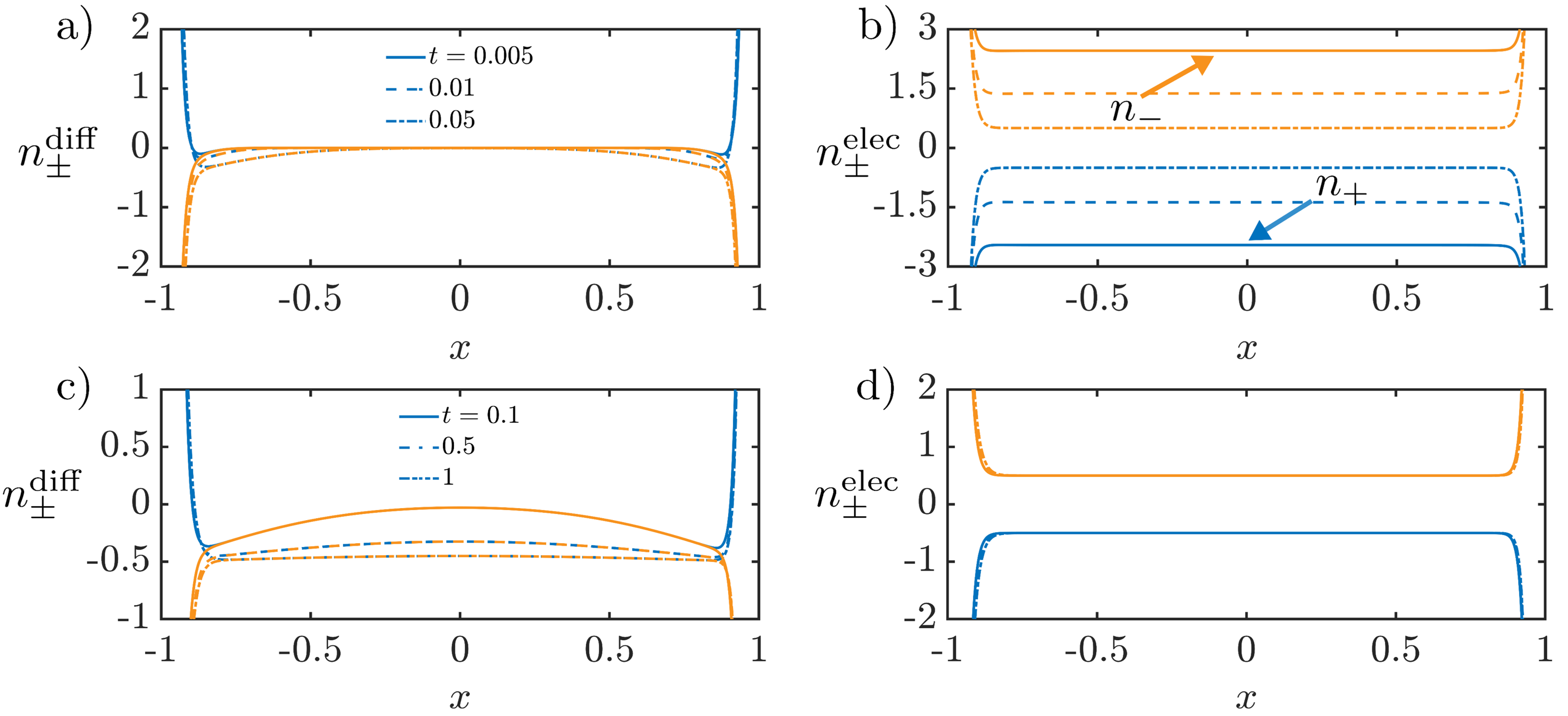}
    \caption{\textbf{Diffusive and electromigrative fluxes at various times.} We define diffusive and electromigrative fluxes as $n_{\pm}^{\textrm{diff}} = - \frac{\partial c_{\pm}}{\partial x}$ and $n_{\pm}^{\textrm{elec}} = \mp c_{\pm} \frac{\partial \phi}{\partial x}$. (a) $n_{\pm}^{\textrm{diff}}$ vs. $x$ and (b) $n_{\pm}^{\textrm{elec}}$ vs. $x$ at $t = 0.005$ (solid), $0.01$ (dashed), $0.05$ (dash-dot). (c) $n_{\pm}^{\textrm{diff}}$ vs. $x$ and (d) $n_{\pm}^{\textrm{elec}}$ vs. $x$ at $t = 0.1$ (solid), $0.5$ (dashed), $1$ (dash-dot). The positive ion fluxes are blue lines and negative ion fluxes are orange lines. In all cases, $\frac{j_0}{j_\mathrm{lim}} = 0.5$, $\phi_s = 5$, $\epsilon = 0.02$, and $\delta = 1$. At short times, primarily the electromigrative flux in the bulk changes. In contrast, the diffusive flux in the bulk predominantly changes at long times.}
    \label{Fig: fluxes}
\end{figure}

\par{} To mechanistically understand the differences at the two timescales, we plot diffusive and electromigrative fluxes as $n_{\pm}^{\textrm{diff}} = - \frac{\partial c_{\pm}}{\partial x}$ and $n_{\pm}^{\textrm{elec}}= \mp c_{\pm} \frac{\partial \phi}{\partial x}$, respectively. Fig. \ref{Fig: fluxes} describes the dependence of scaled fluxes at short (a, b) and long times (c, d); also see supplementary video 2. At both timescales, in the EDL regions, the fluxes are orders of magnitude higher than the bulk, such that $n_{\pm}^{\textrm{diff}}$ is balanced by an equal and opposite $n_{\pm}^{\textrm{elec}}$, which forms the basis of the Boltzmann distribution. At the short timescale, transport in the bulk is governed largely by electromigration, such that the electromigrative flux decreases with time. In contrast, at the long timescale, both diffusion and electromigration control the transport in the bulk. However, the electromigrative flux remains roughly constant whereas the diffusive flux varies both spatially and in time. At the steady state, diffusive fluxes adjust themselves in a way that the combined diffusive and electromigrative flux equals the surface reaction flux. 

\par A similar analysis can be performed for $j_0$ values which depend on concentration and potential values, though additional parameters corresponding to kinetic models may be required. As an example, the usage of a Frumkin-Bultler-Volmer boundary condition would require values of forward and reverse (if relevant) rate constants as well as effective transfer coefficients for the cation and anion. With those additional parameters, one can write $j_0 = k_f c|_{\tilde{x} = 0} \exp\left[\alpha_c \Delta \phi_s \right] - k_r \exp\left[-\alpha_a \Delta \phi_s \right]$ for the left EDL region, where $k_f$ and $k_r$ are the forward and reverse rate constants, respectively, $\alpha_c$ and $\alpha_a$ are the effective transfer coefficients for the cation and anion, respectively, and $\Delta \phi_s$ is the potential drop across the left Stern layer. Note that $\Delta \phi_s$ has a different representation at each timescale, and as such, the form of this boundary condition changes at each timescale. Further, note that any ion concentration dependence of the reverse reaction is said to be implicit within $k_r$ in this example, similar to the work of Biesheuvel et. al. \cite{biesheuvel2011diffuse}. While we do not discuss results of such simulations here for simplicity, once boundary conditions are written in the form of concentrations, potentials, and dimensionless constants, they can be directly applied within the system of equations presented in this work.

\section{Three ions with a constant flux} 
\label{sec: 3ion}
To demonstrate the universality of our approach, we increase the complexity of our system and consider a scenario with three ions. Their concentrations are denoted as $c_1$, $c_2$, and $c_3$ and valences as $z_1=1$, $z_2=-2$, and $z_3=1$. We assume that at $t=0$, $c_1(x,0)=c_{10}=1$, $c_2(x,0)=c_{20}=3/4$, and $c_3(x,0)=c_{30}=1/2$ such that the system is initially electroneutral everywhere, or $\sum_i z_i c_{i0}=0$. For simplicity, we consider only the first ion to react at the electrode surfaces with a constant rate $n_{1l}=n_{1r}=j_0$. Since the second and third ions do not react, $n_{2l}=n_{2r}=0$ and $n_{3l}=n_{3r}=0$. Therefore, $j_l = j_r=j_0$. While it is possible to solve the problem and obtain a time-dependent solution, for brevity, we only evaluate the concentration and potential profiles at the limits $t_{\lambda} \rightarrow \infty$ and $t_D \rightarrow \infty$ for a given $\phi_s$, $\delta$, $\epsilon$ and $j_0$. We note that $j_0$ needs to be less than $j_{\textrm{lim}}$, a discussion for which is provided below. 

\par{} We now describe the calculations for $t_{\lambda} \rightarrow \infty$. Since $j_l=j_0$,  Eq. (\ref{Eq: tilde-curr-eqn-final}) enables us to write
\begin{equation}
    A(\infty) = -\frac{j_0}{\sum z_i^2 c_{i0}}.
    \label{Eq: 3ion-A}
\end{equation}
Further, it is straightforward to subtract Eqs. (\ref{Eq: tilde-curr-eqn-final}) and (\ref{Eq: hat-curr-eqn-final}) to obtain $\tilde{q} (t_\lambda) + \hat{q} (t_\lambda) =0$. Utilizing $\tilde{q} + \hat{q} =0$ and subtracting Eqs. (\ref{Eq: tilde-stern-bc-final}) and (\ref{Eq: hat-stern-bc-final}) yields
\begin{equation}
    \tilde{\gamma} (t_\lambda)+ \hat{\gamma} (t_\lambda) + 2B (t_\lambda) = 0.
    \label{Eq: 3ion-B}
\end{equation}
Eqs. (\ref{Eq: 3ion-A}), (\ref{Eq: 3ion-B}), (\ref{Eq: tilde-stern-bc-final}), and (\ref{Eq: hat-stern-bc-final}) yield a closed system of equations to evaluate $A(\infty)$, $B(\infty)$, $\tilde{\gamma}(\infty)$, and $\hat{\gamma} (\infty)$. To obtain $\tilde{\phi}(\tilde{x}, \infty)$ and $\hat{\phi}(\hat{x}, \infty)$, we integrate Eqs. (\ref{Eq: dphitildedx}) and (\ref{Eq: dphihatdx}) with boundary conditions $\tilde{\phi}(0, \infty) = \tilde{\gamma}(\infty) -A (\infty)+ B (\infty)$ and $\hat{\phi}(0, \infty) = \hat{\gamma}(\infty) + A (\infty) + B (\infty)$, respectively. We then utilize Eq. (\ref{Eq: lamda-full-sol}) to find a spatially composite solution. 
\par
\begin{figure}[h!]
    \centering
    \includegraphics[width=0.95 \textwidth,keepaspectratio]{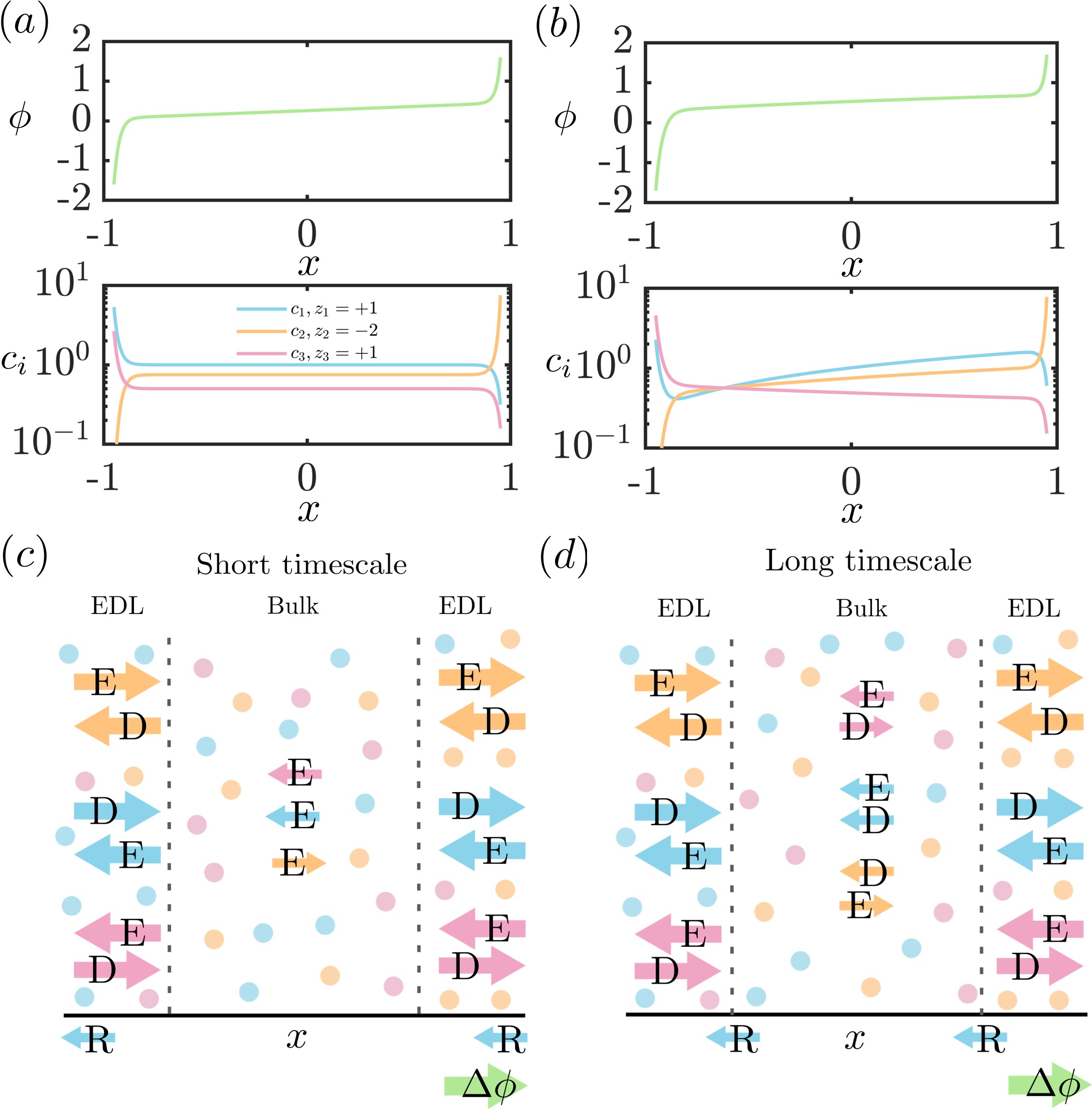}
    \caption{\textbf{Overview of concentration and potential profiles for a three ion system.} (a) potential (green) and concentration (blue, orange, pink) profiles at $t_\lambda \rightarrow \infty$. Ion valences are $+1$ (blue), $-2$ (orange), and $+1$ (pink), with only the blue ion reacting near the electrode surfaces. (b) potential (green) and concentration (blue, orange, pink) profiles at $t_D \rightarrow \infty$. $\phi_s = 5$, $\delta = 1$, $\epsilon = 0.05$, and $j_0 = 0.75 j_{\textrm{lim}}$. (c) Direction of diffusive (denoted as D) and electromigrative (denoted as E) fluxes for each ion at the $t_\lambda$ timescale. Accumulation occurs in the double layers at this timescale, directly influenced by the surface reactive flux of the blue ion, denoted here as R. Electromigration in the bulk is driven by the potential gradient across that region. Diffusion is negligible in the bulk region. (d) Direction of diffusive and electromigrative fluxes for each ion at the $t_D$ timescale. Accumulation occurs in the bulk at this timescale and the flux condition moves from the Stern-diffuse layer interface to the diffuse-bulk layer interface. At steady state, electromigration and diffusion equal the flux that arises due to surface reactions. In both cases, $\nabla \phi$ denotes the direction of the potential gradient.}
    \label{Fig: transport}
\end{figure}

{}Fig. \ref{Fig: transport}(a) provides potential (top panel) and concentration (bottom panel) profiles at $t_{\lambda} \rightarrow \infty$ for $\phi_s=5$, $\epsilon=0.05$, $\delta=1$ and $\frac{j_0}{j_{\textrm{lim}}}=0.75$, where $j_{\textrm{lim}}=-1.197$ (see discussion below Eq. (\ref{Eq: 3ion-c1bar})). The obtained potential profile shows that $B(\infty) \neq 0$, even though the charges stored inside the left and right EDLs are equal and opposite, in contrast to the binary symmetric electrolyte where $B=0$. This difference in potential occurs because asymmetry in valence and initial concentrations for cations and anions leads to different magnitudes of $\tilde{\gamma} (\infty)$ and $\hat{\gamma} (\infty)$. Fig. \ref{Fig: transport}(a) shows that the cations, i.e., the first and third ions, possess a higher concentration at the left EDL whereas the anion has a higher concentration at the right EDL. Importantly, we find that since the valence and initial concentrations of cations and anions are different, the effective Debye lengths differ even at the $t_{\lambda}$ scale, unlike the scenario of a symmetric binary electrolyte. 
\par{} Next, we discuss the solution for $t_{D} \rightarrow \infty$. At the steady state of the long timescale, Eqs. (\ref{Eq: bar-ci-tD}), (\ref{Eq: bar-ni-tD}), and (\ref{Eq: tilde-ni1-tD}) are combined to obtain 
\begin{subequations}
\label{Eq: 3ion-long-tD}
\begin{eqnarray}
- \frac{\partial \bar{c}_1}{\partial \bar{x}} -  \bar{c}_1 \frac{\partial \bar{\phi}}{\partial \bar{x}} &= j_0, \\
- \frac{\partial \bar{c}_2}{\partial \bar{x}} + 2 \bar{c}_2 \frac{\partial \bar{\phi}}{\partial \bar{x}} &= 0, \\
- \frac{\partial \bar{c}_3}{\partial \bar{x}} -  \bar{c}_3 \frac{\partial \bar{\phi}}{\partial \bar{x}} &= 0 \label{Eq: 3ion-dc3dx},  
\end{eqnarray}
\end{subequations}
\noindent Adding all of the Eqs. in (\ref{Eq: 3ion-long-tD}) and employing $\bar{c}_1 + \bar{c}_3 = 2 \bar{c}_2$ reveals
\begin{equation}
    \label{Eq: 3ion-c2bar}
    \bar{c}_2 (\bar{x}, \infty)= \frac{- j_0 \bar{x} + \sum_i c_{i0}}{3},  
\end{equation}
where we have utilized $\int_{-1+\epsilon \delta}^{1-\epsilon \delta} \left( \bar{c}_{i} - c_{i0} \right) d \bar{x}=0 $. Multiplying each Eq. in (\ref{Eq: 3ion-long-tD}) with $z_i$ and adding yields 
\begin{equation}
        \label{Eq: 3ion-dphidx}
    \frac{\partial \bar{\phi}}{\partial \bar{x}} = -\frac{j_0}{6 \bar{c}_2}. 
\end{equation}

Substituting Eq. (\ref{Eq: 3ion-dphidx}) in Eq. (\ref{Eq: 3ion-dc3dx}) enables us to obtain
\begin{equation}
    \label{Eq: 3ion-c3bar}
    \bar{c}_3 (\bar{x}, \infty) = \frac{c_{30}}{2} \frac{\sqrt{\bar{c}_2(1-\epsilon \delta, \infty)} + \sqrt{\bar{c}_2(-1+ \epsilon \delta, \infty)}}{\sqrt{\bar{c}_2(\bar{x}, \infty)}}.
\end{equation}
Finally, Eqs. (\ref{Eq: 3ion-c2bar}) and (\ref{Eq: 3ion-c3bar}) are combined with the electroneutrality requirement to write
\begin{equation}
     \label{Eq: 3ion-c1bar}
    \bar{c}_1 (\bar{x}, \infty) =  \frac{4 \bar{c}_2(\bar{x}, \infty)^{3/2} - c_{30} (\sqrt{\bar{c}_2(1- \epsilon \delta, \infty)} + \sqrt{\bar{c}_2(-1+ \epsilon \delta, \infty)})}{2 \sqrt{\bar{c}_2(\bar{x}, \infty)}}. 
\end{equation}

Eq. (\ref{Eq: 3ion-c3bar}) can be used to ensure $\bar{c}_1 \ge 0$, which gives us the condition for $j_{\textrm{lim}}$ as 
\begin{equation}
   j_{\textrm{lim}} = -\sum_i c_{i0} + \left[ \frac{3 c_{30}}{4} \left( \sqrt{-j_{\textrm{lim}} + \sum_i c_{i0}} + \sqrt{j_{\textrm{lim}} + \sum_i c_{i0}} \right) \right]^{2/3}.
   \label{Eq: 3ion-jlim}
\end{equation}
To estimate $\bar{\phi}(\bar{x}, \infty)$, we integrate Eq. (\ref{Eq: 3ion-dphidx}) with Eqs. (\ref{Eq: tilde-stern-bc}), (\ref{Eq: tilde-q-tD}), (\ref{Eq: hat-q-tD}), (\ref{Eq: hat-stern-bc}), and (\ref{Eq: phi-tilde-hat}) to obtain $\bar{\phi}(\bar{x}, \infty)$, $\tilde{q}(\infty)$, $\hat{q}(\infty)$, $\tilde{\phi}(0, \infty)$, and $\hat{\phi}(0, \infty)$. Next, $\tilde{\phi}(\tilde{x}, \infty)$ and $\hat{\phi}(\hat{x}, \infty)$ are obtained by numerically integrating Eqs. (\ref{Eq: dphitildedx}) and (\ref{Eq: dphihatdx}) with the obtained values of $\tilde{\phi}(0, \infty)$ and $\hat{\phi}(0, \infty)$ utilized as boundary conditions. Finally, a composite solution is obtained by using Eqs. (\ref{Eq: D-full-sol}).
\par{}Fig. \ref{Fig: transport}(b) provides potential (top panel) and concentration (bottom panel) profiles for conditions identical to panel (a). Similar to the scenario of the binary symmetric electrolyte, we find that a mild curvature starts to appear in the potential profile in the bulk region. More significantly, the concentration profiles of ions develop significant gradients in the bulk region. Interestingly, while the first and second ions have a positive concentration gradient in the bulk region, the third ion has a negative concentration gradient. We elaborate on this effect below.

\par{} The concentration and potential profiles are set by the combined effects of the diffusive and electromigrative fluxes with the surface reaction flux. These combinations differ significantly at the two timescales, as shown in Fig. \ref{Fig: transport}(c),(d). We initially focus on the transport processes at the EDL charging timescale (Fig. \ref{Fig: transport}(c)). Since at the EDL timescale, the concentration gradients are limited to the EDLs, the concentrations of ions remain constant (see Eq. (\ref{Eq: ci0})) and the diffusive flux is negligible in the bulk. In contrast, the current leaking from the EDLs requires a potential gradient in the bulk, which implies that electromigration is still present in the bulk. The direction of the electromigration flux is set by the charge of a given ion. Therefore, the cations have an electromigrative flux towards negative $x$ whereas the anion has an electromigrative flux towards positive $x$. Inside the EDLs, the Boltzmann distribution still applies (see Eqs. (\ref{Eq: tilde-PB}) and (\ref{Eq: hat-PB})). Therefore, a balance between diffusion and electromigration fluxes is required. Moreover, due to the thinness of the EDLs, both of the fluxes are significantly larger (see Fig. \ref{Fig: fluxes}) and the direction of the fluxes is set by the charge of the ion.  Since the EDLs are getting charged, an accumulation term in the double layers is required for the EDLs to develop. This accumulation is directly, but not exclusively, dependent on the surface reaction flux $j_0$. We emphasize that surface reactions directly impact the fluxes at this timescale and their effects should not be ignored at the $t_\lambda$ scale.
\par{} Next, we discuss the transport processses at the bulk diffusion timescale, as displayed in Fig. \ref{Fig: transport}(d). At this timescale, the EDLs are fully charged and can no longer accumulate additional charge. Therefore, the flux boundary condition is transferred from the Stern-diffuse layer interface to the diffuse-bulk interface. Since the Boltzmann distribution is required to hold at this timescale, albeit with a different reference point (see Eqs. (\ref{Eq: tilde-PB-tD}) and (\ref{Eq: hat-PB-tD})), large diffusion and electromigration fluxes are still present in both EDLs and the directions of the fluxes are identical to the $t_{\lambda}$ scale. As a consequence of the surface boundary condition moving to the diffuse-bulk interface, accumulation and concentration gradients start to appear in the bulk. Therefore, a diffusive flux begins to appear in the bulk region. We note that the direction of electromigration is identical to the $t_{\lambda}$ scale and the concentration gradients develop to equal the surface reaction flux. For instance, with the first ion, both electromigration and diffusion act in the negative $x$ direction and together equal the surface reaction flux, as this is the only reactive ion in this system. In contrast, the second ion has electomigrative flux in the positive $x$ direction and diffusive flux in the negative $x$ direction in order to keep a net zero flux at steady state. The third ion is completely opposite to the second ion, i.e., an electromigrative flux in the negative $x$ direction and a diffusive flux in the positive $x$ direction. Finally, while the direction of the electromigration flux remains identical to that of the shorter timescale, quantitatively, there is a difference since the local ion concentrations need to satisfy electroneutrality (see Fig. \ref{Fig: transport}(b)).  As is evident from this discussion, the relationship between electromigration and diffusion is non-trivial, and is significantly dependent on the combination of ions and electrochemical reactions. 

\section{Conclusions}
\label{sec: conclusions}
In this article, we lay out a theoretical framework that is computationally efficient and is able to simulate electrochemical cells with an arbitrary number of ionic components while accounting for the coupled effects of EDLs and redox reactions. The theoretical framework presented in this paper enables us to unravel the physics for multicomponent systems with minimal computational requirements and can be adapted to practical scenarios where multicomponent electrolytes are prevalent. Specifically, in this article we: \vspace{-0.5cm}
\begin{enumerate}
    \item Formulate a perturbation model that couples double-layer charging with electrochemical reactions near flat plate electrode surfaces for an arbitrary number of ions and any type of kinetic model. \vspace{-0.5cm}
    \item Employ asymptotic matching to stitch six solutions (three spatial regions at two timescales) to form one composite solution, as summarized in Table \ref{tab:general}. The approach outlined in Table \ref{tab:general} assumes EDLs are thin and that ion diffusivities are equal, but imposes no other restrictions. \vspace{-0.5cm}
    \item Validate our model by achieving excellent agreement with direct numerical simulations. \vspace{-0.5cm}
    \item Demonstrate that the proposed approach remains numerically stable for large potentials up to approximately 0.5-1 V. \vspace{-0.5cm}
    \item Underscore that surface reactions have a direct effect on double-layer charging and the capacitance of EDLs by analyzing the dynamics of a binary symmetric electrolyte and a three-ion system.\vspace{-0.5cm}
    \item Provide physical insights into the relationships between transport processes in each spatial region, which differ at short and long times.
\end{enumerate} 
\vspace{-0.5cm}
We note that while the model here assumes point charges, it is straightforward to extend to finite ion sizes through the addition of additional factors to the Poisson-Nernst-Planck equations \cite{kilic2007steric, kilic2007stericpartone}. In the future, we also wish to extend our analysis to unequal diffusivities, which we believe would require asymptotic expansions of ion diffusivities in order to eliminate difficulties with asymptotic formulation. Our analysis can also be extended to incorporate volumetric electrochemical reactions through the presence of a reaction term and a Damk\"{o}hler number in the Nernst-Planck equations. The perturbation model would then have three timescales instead of two: one each for double layer formation, bulk diffusion, and reaction kinetics, all of which can be decomposed and matched to form a composite solution assuming the reaction kinetics-dominated timescale is not on the same order of either of the other two, which would cause asymptotic ordering to fail. We believe that the approach outlined in this article is an important step towards accurate simulations of multicomponent electrochemical systems, which are becoming increasingly important for energy storage and sustainability applications \cite{zhou2021electrolytes,jung2017oxygen,crothers2020theory,zhong2015review,ZHANG2018314}. 

\section*{Conflicts of Interest}
There are no conflicts to declare.
\section*{Acknowledgements}
The authors would like to thank Adam Holewinski, Ali Mani, Howard Stone, Paige Brimley, and Robert Davis for their helpful input regarding this project. N. J. and A. G. would also like to acknowledge the financial support from the GAANN program in Polymer Materials for Energy and Sustainability. F. H. and A. G. would like to thank the Ryland Family Graduate Fellowship for financial support.

\pagebreak

\section*{Nomenclature}
\begin{table}[h!]
\caption{\textbf{Table of nomenclature.} We note that all variables with bar, tilde, and hat correspond to the bulk electrolyte, left-double-layer, and right-double-layer regions, respectively.}
\resizebox{0.95\textwidth}{!}{
\begin{tabular}{|l|l|} 
\hline
 \textbf{Dimensional Nomenclature} & \textbf{Dimensionless Nomenclature} \\ 
 \hline
$C_i$ \hspace{0.25cm}concentration of $i^{\textrm{th}}$ ion & $c_i$ \hspace{0.25cm}concentration of $i^{\textrm{th}}$ ion  \\ 
$C_{i0}$ \hspace{0.15cm}initial concentration of $i^{\textrm{th}}$ ion &  $c_{i0}$ \hspace{0.1cm}initial concentration of $i^{\textrm{th}}$ ion \\ 
$\tau$ \hspace{0.3cm}  time & $t$ \hspace{0.25cm}  time \\ 
$N_i$ \hspace{0.2cm}flux of $i^{\textrm{th}}$ ion & $n_i$ \hspace{0.2cm}flux of $i^{\textrm{th}}$ ion\\ 
$L$ \hspace{0.25cm}  boundary value of the spatial coordinate $X$ & $\epsilon$ \hspace{0.25cm}   ratio of diffuse layer length to total system length, or $\epsilon=\frac{\lambda}{L}$ \\ 
$X$ \hspace{0.25cm}  spatial coordinate, with bounds $-L+\lambda_s \le X \le L-\lambda_s$  & $x$ \hspace{0.25cm}  spatial coordinate, with bounds $-1 \le x \le 1$  \\ 
$\mathcal{D}$ \hspace{0.25cm}  ion diffusivities, assumed to be equal &  $\bar{x}$ \hspace{0.25cm}  spatial coordinate for bulk electrolyte region, \\ $C^*$ \hspace{0.25cm}reference concentration, used for nondimensionalization& or $\bar{x} = x$ with bounds $-1+ \epsilon \delta \le \bar{x} \le 1- \epsilon \delta$  \\
$z_i$ \hspace{0.25cm} valence of $i^{\textrm{th}}$ ion & $\tilde{x}$ \hspace{0.25cm}  spatial coordinate for left-double-layer region, \\$\lambda$ \hspace{0.3cm} measure of characteristic diffuse layer length, $\lambda= \sqrt{ \frac{\varepsilon k_B T}{e^2 C^*}}$ & or $\tilde{x} = \frac{x+1 - \epsilon \delta }{\epsilon}$ with bounds $ 0 \le \tilde{x} \le \infty$ \\
$e$ \hspace{0.35cm} charge of an electron & $\hat{x}$ \hspace{0.25cm} spatial coordinate for right-double-layer region, \\& or $\hat{x} = \frac{-x + 1 - \epsilon \delta}{\epsilon}$ with bounds $0 \le \hat{x} \le \infty$ \\
$k_B$ \hspace{0.2cm}Boltzmann constant & $t_\lambda$ \hspace{0.25cm}time variable at the short (double-layer charging) timescale \\ 
$T$ \hspace{0.25cm} temperature & $t_D$ \hspace{0.2cm}time variable at the long (bulk diffusion) timescale \\ 
$\Phi$ \hspace{0.35cm}electric potential & $\phi$ \hspace{0.35cm}electric potential  \\ 
$\Phi_s$ \hspace{0.2cm}magnitude of the applied potential at the boundaries & $\phi_s$ \hspace{0.2cm}magnitude of the applied potential at the boundaries \\
$\varepsilon$ \hspace{0.42cm}electrical permittivity &  $q$ \hspace{0.4cm}total charge stored inside the diffuse layer  \\ 
$Q_e$ \hspace{0.15cm}volumetric charge density, or $Q_e= \sum_i z_i e C_i$ & $\rho_e$ \hspace{0.2cm}volumetric charge density, or $\rho_e = \sum z_i c_i $  \\
$\lambda_s$ \hspace{0.25cm}length of the Stern layer  &  $\delta$ \hspace{0.25cm} ratio of Stern layer length to diffuse layer length, or $\delta=\frac{\lambda_s}{\lambda}$  \\ 
$N_{il}$ \hspace{0.15cm}flux of $i^{\textrm{th}}$ ion at the left boundary & $n_{il}$ \hspace{0.15cm}flux of $i^{\textrm{th}}$ ion at the left boundary  \\
$N_{ir}$ \hspace{0.1cm}flux of $i^{\textrm{th}}$ ion at the right boundary & $n_{ir}$ \hspace{0.1cm}flux of $i^{\textrm{th}}$ ion at the right boundary  \\
\hline
\end{tabular}
}
\label{tab:nomenclature}
\end{table}

\def\bibsection{\section*{\refname}}
\bibliographystyle{ieeetr}
\bibliography{apssample}

\end{document}